\documentclass[journal,usletter]{IEEEtran/IEEEtran}
\usepackage[utf8]{inputenc}
\usepackage[T1]{fontenc}
\usepackage{siunitx}
\usepackage{hyperref}
\usepackage{cite}
\usepackage[frozencache=true]{minted}
\usepackage{adjustbox}
\usepackage{algorithmic}
\usepackage{graphicx}
\usepackage{textcomp}
\usepackage{makecell}
\usepackage{xcolor}
\usepackage{color}
\usepackage{multicol}
\usepackage{caption}
\usepackage{ifthen}
\usepackage{soulutf8} 
\usepackage{float}
\usepackage{subcaption}
\usepackage{enumitem}
\usepackage{soul}
\usepackage[absolute]{textpos}
\usepackage{mathtools}
\usepackage{tcolorbox}
\tcbuselibrary{skins}
\usepackage{tikz}
\usetikzlibrary{shapes}
\usetikzlibrary{patterns, fit}
\pgfdeclarelayer{bg}    
\pgfsetlayers{bg,main}  
\usepackage{pdfpages}

\setminted[java]{
xleftmargin=20pt,
framesep=2mm,
linenos=true,
breaklines=true,
tabsize=4,
encoding=utf8,
fontsize=\footnotesize,
frame=none,
escapeinside=||
}

\definecolor{hlred}{RGB}{255, 204, 204}

\newcommand{\ts}[0]{\textsc{TriggerScope}}
\newcommand{\tso}[0]{\textsc{TSOpen}}

\newcommand{\dataset}[0]{\textsc{TrigDB}}

\newcommand{\myconclusion}[1]{
\begin{center}
\begin{tikzpicture}
\node[draw, rounded corners, fill=black!10, inner sep=.25cm] {
\begin{minipage}{0.92\columnwidth}
#1
\end{minipage}
};
\end{tikzpicture}
\end{center}
}

\newcommand{\myshortconclusion}[1]{
\begin{center}
\begin{tikzpicture}
\node[draw, rounded corners, fill=white, inner sep=.25cm] {
\begin{minipage}{0.92\columnwidth}
#1
\end{minipage}
};
\end{tikzpicture}
\end{center}
}

\begin{document}

\title{On The (In)Effectiveness of Static Logic Bomb Detection for Android Apps}

\author{
Jordan~Samhi$^{1}$ Alexandre~Bartel$^{1,2,3,*}$
\thanks{$^{*}$At the time this research was conducted 
Alexandre Bartel was at the University of Luxembourg and
the University of Copenhagen.}\\
$^{1}$~\IEEEmembership{University of Luxembourg} 
$^{2}$~\IEEEmembership{University of Copenhagen}
$^{3}$~\IEEEmembership{Umeå University}
}

\maketitle

\begin{abstract}
Android is present in more than 85\% of mobile devices, making it a prime target for malware. 
Malicious code is becoming increasingly sophisticated and relies on \emph{logic bombs} to hide
itself from dynamic analysis.
In this paper, we perform a large scale study of \tso{}, our open-source implementation of
the state-of-the-art static logic bomb scanner \ts{}, on more than 500k Android applications. Results indicate that the approach scales.
Moreover, we investigate the discrepancies and show that the approach can reach a very low false-positive rate, 0.3\%, but at a particular cost, e.g., removing 90\% of sensitive methods. 
Therefore, it might not be realistic to rely on such an approach to automatically detect \emph{all} logic bombs in large datasets.
However, it could be used to speed up the location of malicious code, for instance, while reverse engineering applications.
We also present \dataset{} a database of 68 Android applications containing trigger-based behavior as a ground-truth to the research community.

\end{abstract}


\section{Introduction}
\label{introduction}
\label{different_types_lb}
Android is the most popular mobile operating system with more than 85\% of the market share in 2020~\cite{idc}, which undeniably makes it a target of choice for attackers.
Fortunately, Google set up different solutions to secure access for applications in their \emph{Google Play}.
It ranges from fully-automated programs using state-of-the-art technologies (e.g., Google Play Protect~\cite{gpp}) to manual reviews of randomly selected applications.
The predominant opinion is that the \emph{Google Play} market is considered relatively malware-free.
However, as automated techniques are not entirely reliable and manually reviewing every submitted application is not possible, they continuously improve their solutions' precision and continue to analyze already-present-in-store applications.

Consequently, the main challenge for attackers is to build malicious applications that remain under the radar of automated techniques.
For this purpose, they can obfuscate the code to make the analysis more difficult.
Example of obfuscation includes code manipulation techniques~\cite{chan2004advanced}, use of dynamic code loading~\cite{liang1998dynamic} or use of the Java reflection API~\cite{huang2014type}.
Attackers can also use other techniques such as packing~\cite{zhang2015dexhunter} 
which relies on encryption to hide their malicious code.
In this paper, we focus on one type of evasion technique based on logic bombs.
A logic bomb is code logic which executes malicious code only when particular conditions are met.

A classic example would be malicious code triggered only if the application is not running in a sandboxed environment or after a hard-coded date, making it invisible for dynamic analyses.
This behavior shows how simple code logic can defeat most dynamic analyses leading to undetected malicious applications.

In the last decade, researchers have developed multiple tools
to help detecting logic bombs~\cite{brumley2008automatically}~\cite{zheng2012smartdroid}~\cite{papp2017towards}.
Most of them are either 
not fully automated,
not generic or 
have a low recall.
However, one approach, \ts{}~\cite{fratantonio2016triggerscope} stands out because 
it is fully automated and has a false positive rate close to 0.3\%.
In this paper, we try to replicate \ts{} and perform a large-scale study to show that the approach scales.
Furthermore, we identify specific parameters that have a direct impact on the false positive rate.

As \ts{} is not publicly available and the authors cannot share the tool,
we implement their approach as an open-source version called \tso{}.
Although \tso{} has been implemented by faithfully following the details of the approach given in \ts{} paper, 
we did not use the same programming language, i.e., C++.
We used Java to reuse publicly available and well tested state-of-the-art solutions.
Indeed, our solution relies on the so-called Soot framework~\cite{vallee2010soot} to convert the Java bytecode into an intermediate representation called Jimple~\cite{vallee1998jimple} and to automatically construct control flow graphs.
Also, to model the Android framework, the life-cycle of each component and the inter-component communication, 
\tso{} relies on algorithms from FlowDroid~\cite{arzt2014flowdroid}.

We use \tso{} to conduct a large-scale analysis to see if such a static approach is scalable.
We ran \tso{} over a set of \num{508122} applications from a well-known database of Android applications named \textsc{Androzoo}~\cite{allix2016androzoo}.
This experiment shows that the approach is scalable but yields a high false-positive rate.
Hence, because of this high false-positive rate, the approach might not be suitable to detect \emph{all} logic bombs automatically.
More than \num{99651} applications were flagged with \num{522300} triggers supposedly malicious, yielding a false-positive rate of more than 17\%.

Since we obtain a false-positive rate which is much higher than in the literature, we investigate the discrepancies.
We construct multiple datasets to consider the concept drift effect, which could affect the results shown by Jordaney et al.~\cite{jordaney2017transcend}. 
Moreover, we also investigate multiple aspects of the implementation, such as the list of sensitive methods, 
the call-graph construction algorithm or
the timeout threshold. 
Furthermore, we applied additional two filters not mentioned in the literature: 
(1) Purely symbolic values removal and 
(2) Different package name removal.
Results indicate that to get close to a false positive rate of 0.3\%, 
either aggressive filters should be put in place or 
a short list of sensitive methods should be used.
In both cases, the impact on the false-negative rate is considerable.
This means that if the approach is usable in practice with a low false-positive rate, it might miss many applications containing logic bombs.

We have sent the paper to the \ts{}'s authors, who gave us positive feedback and did not see any significant issue regarding our approach nor on \tso{}'s design.

In summary, we present the following contributions :
\begin{itemize}
\item We implement \tso{}, the first open-source version of the state-of-the-art approach for detecting logic bombs and show that this approach might not be appropriate for \textbf{automatically} detecting logic bombs because it yields too many false-positives.
\item We conduct a large-scale analysis over a set of more than \num{500000} Android applications.

While the approach is theoretically not scalable because it relies on NP-hard algorithms, 
we find that, in practice, 80\% of the applications can be analyzed.
\item We conduct multiple experiments on the approach's parameters to see the impact on the false positive rate and identify that a low false-positive rate can be reached but at the expense, for instance, of missing a large number of sensitive methods.
\item We experimentally show that \ts{}'s approach might not be usable in a realistic setting to detect logic bombs with the information given in the original paper. We empirically show that using \ts{}'s approach, \emph{trigger analysis} is not sufficient to detect logic bombs.
\item We publicly release a database of Android apps containing logic bombs as a ground-truth for future research.
\end{itemize}

We make available our implementation of \tso{} and \dataset{} with datasets to reproduce our experimental results:

\begin{center}
\url{https://github.com/JordanSamhi/TSOpen}
\end{center}

The remainder of the paper is organized as follows.
First, a motivation example is given in Section~\ref{motivation} in order to clarify the reason we are studying logic bombs.
Afterward comes the overview of \ts{} in Section~\ref{overview}, in which we give an overview of its structure.
We evaluate \tso{} in Section~\ref{evaluation}.
Subsequently, in Section~\ref{limitations}, we discuss the limitations of our work and the reference paper~\cite{fratantonio2016triggerscope}.
Section~\ref{related_work} relates similar state-of-the-art works in the context of detecting anti-reverse-engineering practices.
Finally, in Section~\ref{future_work} we present the direction in which we will continue our research.

\begin{figure*}[ht]
\centering
\input{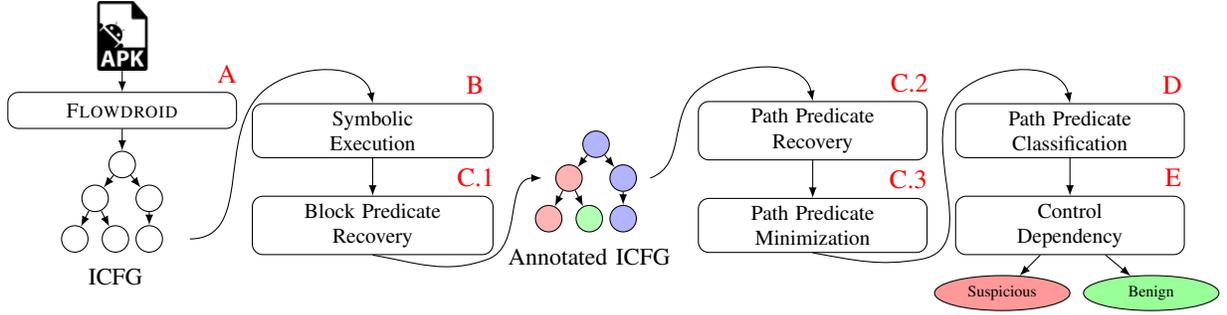}
\caption{Overview TSOpen. First, the application APK is processed by \textsc{Flowdroid} to model the application. Then, every step of the analysis is applied to the ICFG until the final decision of the application's suspiciousness is taken by our tool.}
\label{analysis_overview}
\end{figure*}

\section{Motivation}
\label{motivation}

We consider two motivating examples. 
The first one is an application that seems legit but embeds a time-bomb that is triggered at a specific date:
let's say two weeks after the installation date.
This would mean that the malicious code will remain silent for some time before being triggered. 
Listing~\ref{timeBombExample} is a concrete example of such a logic bomb.
It is located at line 4 within the \texttt{onStart()} method, and it triggers the malicious code when a user opens the application and the date is reached.

We can see that with a minimum of effort, a malware developer can bypass most of the dynamic analyses which study the behavior of applications by monitoring them.
Petsas \& al.~\cite{petsas2014rage} give interesting results regarding simple solutions to bypass state-of-the-art dynamic analyses like Andrubis~\cite{lindorfer2014andrubis} and CopperDroid~\cite{tam2015copperdroid}.
The idea is to keep the malicious code dormant during dynamic analyses by hiding it behind unexplored branches.
Nevertheless, recent works show that dynamic analyses improve code coverage during the execution of an application by forcing path exploration.
They do so by instrumenting the application to modify the control flow.
However, these approaches face several limitations, e.g., \textsc{GroddDroid}~\cite{abraham2015grodddroid} does not force all branches but only those that guard malicious code which they have to detect beforehand, reducing the problem to detecting malicious code.
Likewise, \textsc{X-Force}~\cite{peng2014x} does not force all branches and can force unfeasible paths, making it unsound for analyses.

The first example in Listing~\ref{timeBombExample} was not part of a targeted attack but more in the idea of a widespread malicious application. 
Let's now consider a State-Sponsored Attack or an Advanced Persistent Threat~\cite{chen2014study} which targets specific devices. 
Those devices could embed seemingly legitimate applications that contain, e.g., an SMS-bomb and remain undetected by most of the analysis tools.

\begin{listing}
\begin{minted}{java}
protected void onStart() {
    Date now = new Date();
    Date attackDate = installDate.plusDays(14);
    if(now.after(attackDate)){|\textcolor{red}{attack();}|}
}
\end{minted}
\caption{Time-bomb example}
\label{timeBombExample}
\end{listing}

\begin{listing}
\begin{minted}{java}
public void onReceive(Context c, Intent i) {
	SmsMessage sms = getIncomingSms(i);
	String body = sms.getMessageBody();
	if(body.startsWith("!CMD:"))
        |\textcolor{red}{processCmd(getCmdFromBody(body));}|
}
\end{minted}
\caption{SMS-bomb example}
\label{smsBombExample}
\end{listing}

Such a logic bomb could be used as a backdoor to steal sensitive user data. 
Indeed, if it is a well prepared targeted attack, the attacker could send an SMS with a specific string recognized by the application. 
Then the application would leak wanted information and stop the broadcast of the SMS to other applications supposed to receive it. 
This example shows how important it is to detect as many logic bombs as possible in applications, especially in mobile devices, which nowadays store a lot of personal information.

Listing~\ref{smsBombExample} shows an implementation of such a threat in the method \texttt{onReceive(Context c, Intent i)} of \texttt{BroadcastReceiver} class which is triggered when, in this case, an SMS is received by the device.
First, the body of the SMS is retrieved in line 3. 
Then, at line 4, it is compared against "!CMD:" which is a hardcoded string matched against the body of an SMS.
This check is considered suspicious because the application can match a hardcoded string against any incoming SMS. Hence it can wait for external commands to be executed by the malicious part of the application.
Note that it could be harmless. That is why it is suspicious and not malicious.
If the condition is satisfied, the command is retrieved from the SMS, and the malicious code is activated at line 5.

\section{Overview}
\label{overview}
In this section, we explain \tso{}, an open-source implementation of \ts{}, at the conceptual level. 
The approach is summarized in Figure~\ref{analysis_overview}.

\subsection{Applications representation}
\label{program_representation}
Android apps do not have a single entry point like usual Java programs.
They are made of components, each one having its life-cycle managed by the Android framework.

Modeling life-cycles and how components are connected is not trivial.
That is why \textsc{TSOpen} relies on \textsc{Flowdroid}~\cite{arzt2014flowdroid}.
Indeed, \textsc{Flowdroid} handles intra-component communications by introducing dummy main methods and opaque predicates to guarantee that any execution order would not influence any static analysis over the model.
\textsc{Flowdroid} also relies on \textsc{IccTA}~\cite{li2015iccta} to model the inter-component communications thanks to \textsc{Epicc}~\cite{octeau2013effective}.
Using state-of-the-art solutions allowed us to avoid reimplementing the Android framework modeling
reducing implementation errors.
Thus, we retrieve an \emph{interprocedural control flow graph} on which we can run static analysis algorithms (step A in Figure~\ref{analysis_overview}).

Now that we have a model of the application that can be seen in Appendix~\ref{android_model_flowdroid}, we can run our analysis, starting with the symbolic execution.

\subsection{Symbolic execution}
When classifying predicates, the program has to make decisions depending on the type of objects in conditions, i.e., the condition's semantics.
Therefore, this analysis models, using symbolic execution (step B in Figure~\ref{analysis_overview}), the values and the operations performed over Java objects.
More precisely, as we faithfully implemented \textsc{TriggerScope}, we focus on modeling strings, integers, location, SMS, and time-related objects.
Also, these interesting objects are annotated in order to ease the classification.

Furthermore, the classification cannot be done without retrieving the instructions guarded by a condition, that is why the next step, i.e., the predicate recovery, is essential.

\subsection{Predicate recovery}
An essential step of the analysis is to construct the intra-procedural path predicate related to each instruction to build the logical formula leading to the instruction.
For this, we operate as follows: 

Let $ICFG=(I_r,E_r)$ the directed graph describing the interprocedural control flow graph given by \textsc{Flowdroid} where,
$I_r$ represents the set of reachable instructions of the program and $E_r \subseteq I_r \times I_r$ corresponds to the set of reachable directed edges of the program represented by a pair of instructions $(i_a,i_b)$ indicating that the flow goes from $i_a$ to $i_b$.
Let $C_r$ the set of reachable conditions of the program and $\Gamma^-(i)=\lbrace x \: \vert \: (x,i) \in E_r \rbrace$ the predecessor function.

The algorithm to retrieve the full path predicate of each instruction is described as follows:

\begin{enumerate}
  \item $\forall \: i \in I_r, \forall \: e = \lbrace(x,i) \: \vert \: x \in \Gamma^-(i)\rbrace \in E_r$, annotate $e$ with the closest preceding condition $c \in C_r$ (step C.1 in Figure~\ref{analysis_overview}).
  \item $\forall \: i \in I_r$ annotate $i$ with $p = getFormula(i) = \\ \lbrace \bigvee (getFormula(x) \wedge c) \: \vert \: x \in \Gamma^-(i), \: c$ the condition annotated on edge $ (x,i) \rbrace$ (step C.2 in Figure~\ref{analysis_overview}).
  \item $\forall \: i \in I_r,$ simplify the formula $p$ with the basic laws of Boolean algebra, $p$ is the full intraprocedural path predicate annotated on $i$ (step C.3 in Figure~\ref{analysis_overview}).
\end{enumerate}

The last step is essential in order to remove false dependencies of instructions.
Indeed, consider the following formula: $(p \wedge q) \vee (\neg p \wedge q)$ which could have been calculated after step 2.
The instruction annotated with this formula would have a false dependencies on predicate $p$ because $(p \wedge q) \vee (\neg p \wedge q) = q \wedge (p \vee \neg p) = q \wedge 1 = q$ as defined by the distributive and complementation laws of boolean algebra.
Hence, the elimination of false dependencies.

Now that we have retrieved path predicates and eliminated false dependencies, we can classify predicates.

\subsection{Predicate classification}
In order to classify predicates, i.e., their potential suspiciousness, two essential characteristics are taken into account (step D in Figure~\ref{analysis_overview}).
Firstly, we verify that the predicate involves a previously computed time-, SMS- or location-related object.
Secondly, we verify the type of check performed over the object.
The focus is set to comparisons with relevant previously modeled objects and hardcoded values/constants in the application.
If a condition corresponds to these criteria, it is flagged as suspicious.
Note that the Jimple intermediate representation of the Java bytecode is convenient for this stage as it allows analysts to access explicit object types.
This step acts as a filter for the final control dependency step as it reduces the conditions to analyze by ruling out not suspicious conditions.

We can now perform the last step to check if a sensitive method is called within the guarded instructions of a suspicious condition.

\subsection{Control dependency}
\label{control_dependency}
The last step of the approach consists in characterizing whether a condition is defined as a logic bomb (step E in Figure~\ref{analysis_overview}).
For this, every guarded instruction of a considered condition is checked to verify if it invokes a sensitive method.
Also, \textsc{TriggerScope}'s developers had the idea to check whether a variable would be modified and later involved in another check,
which, in turn, its guarded instructions would be similarly checked.
This idea extends the range of possibilities regarding the search for logic bombs.
Our implementation takes all these steps into account.

The interested reader can find further details about the implementation in Appendix~\ref{implementation}.
Furthermore, Appendix~\ref{example} shows an execution example of the analysis.

\section{Evaluation}
\label{evaluation}

In this section we evaluate \tso{} 
and address the following research questions:

\begin{description}
   
  \item[RQ1:] Does \tso{}'s approach scale?   
  \item[RQ2:] What parameters can impact the false positive rate?
  \item[RQ3:] Is it possible to locate the malicious code with logic bomb detection?
  \item[RQ4:] Do benign and malicious applications use similar behavior regarding the approach under study and why?
  \item[RQ5:] Are \ts{}'s results reproducible?
\end{description}

Our analyses were run on a server with an Intel Xeon E5-2430 2.20GHz processor with 24 cores, and 95GB of RAM and the \emph{High-Performance Computing}~\cite{varrette2014management}
equipment available at the University of Luxembourg.

\subsection{\textbf{RQ1: Does \tso{}'s approach scale?}}
\label{rq_large_scale}
We perform the large scale analysis on a large dataset containing \num{508122} applications.
This dataset has been created by randomly extracting applications from the 10 million applications of Androzoo~\cite{allix2016androzoo}.
This analysis is necessary to understand why it could or could not be deployed in real-world analyses, e.g., before being accepted into a store.
Analyzing millions of applications with a 1-hour timeout has to be parallelized to take as short a time as possible.
For this purpose, we took advantage of the \emph{High-Performance Computing}~\cite{varrette2014management}
equipment available at the University of Luxembourg.

\begin{figure}
\centering
    \includegraphics[scale=0.33]{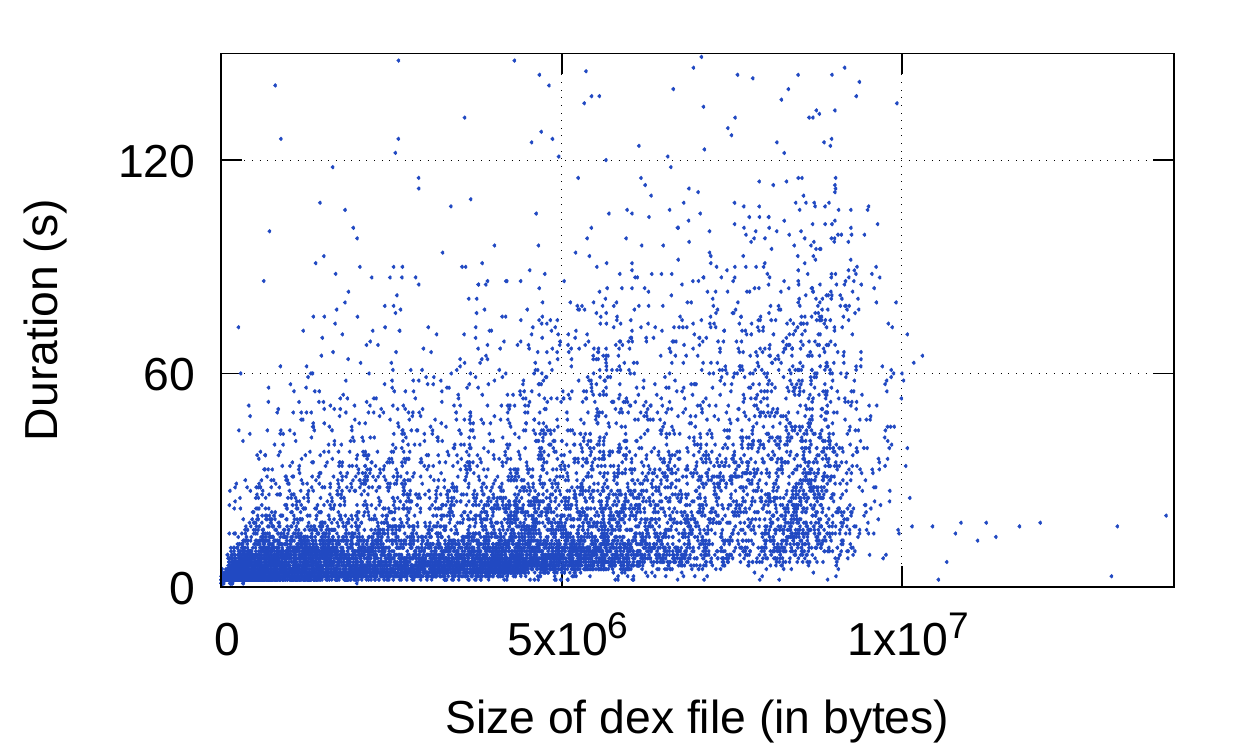}
    \includegraphics[scale=0.33]{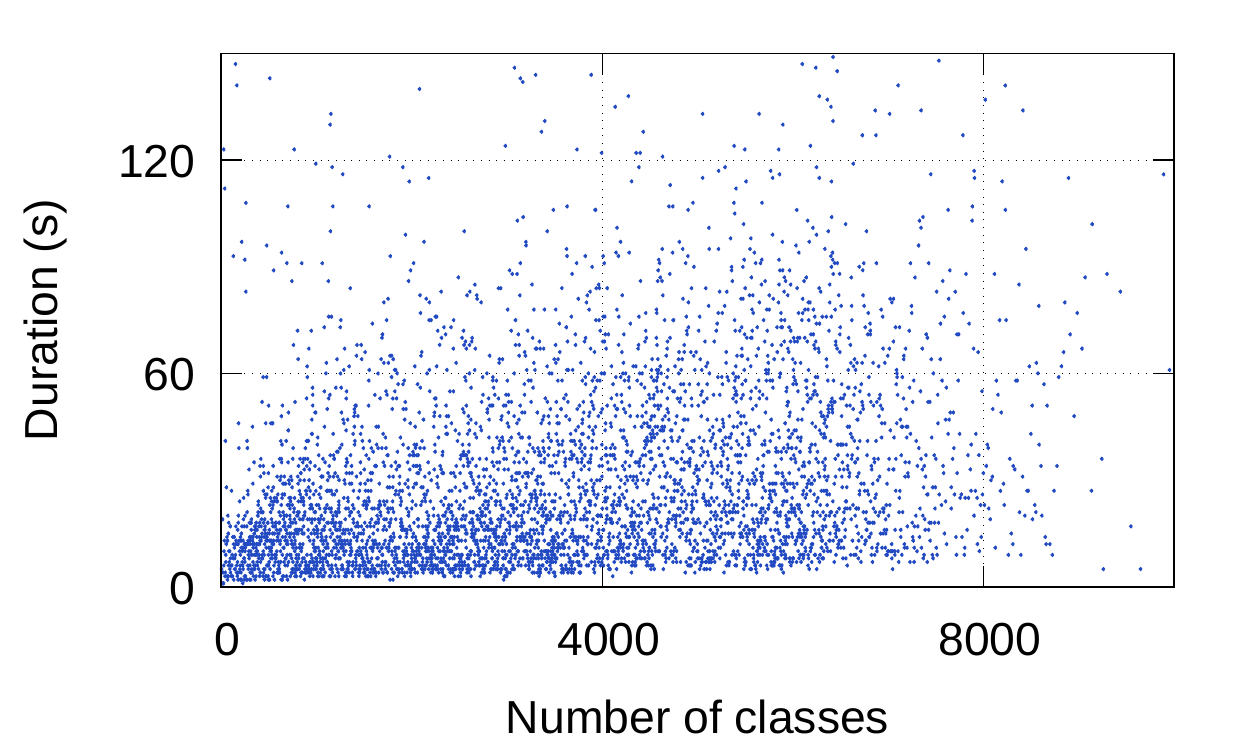}
    \caption{Evolution of the duration of the analysis depending on the size of the dex file and the number of classes in the applications considered.}
  \label{features_time_cloud}
\end{figure}

We took into consideration \num{508122} benign and malicious applications.
In fact, out of \num{508122} considered applications, \num{405810} (\num{79.9}\%) were successfully analyzed with an average of 21 seconds per analysis (e.g., timeout).
The proportion of applications with detected triggers with this approach is \num{19.6}\% (\num{99651}) and \num{34.9}\% (\num{177112}) without library filter (The library filter is further explained in Section~\ref{sec:libraryfilter}).
Also, \num{522300} (\num{791364} without library filter) triggers are detected by \tso{} among which \num{0.48}\% of SMS-related triggers, \num{1.35}\% location-related triggers, and \num{98.17}\% of time-related triggers.

Our default threshold for the timeout is one hour.
Some applications cannot be analyzed within one hour.
A malware developer could simply use techniques, such as obfuscation, to slow down static analysis tools to prevent the application from being analyzed.
To understand how an attacker could bypass the analysis, we measured the execution time of the analysis in function of four features:
(a) the size of dex files, (b) the number of classes, (c) the number of objects, and (d) the number of branches.

In Figure~\ref{features_time_cloud}, we can intuitively assume that there is no correlation between the size of the dex file or the number of classes in the app with the duration of the analysis.
In fact, when measuring the \textit{Pearson Correlation Coefficient (PCC)}, computed based on Equation~\ref{pcc}, we can state that there is no correlation.

\begin{equation}
 r_{xy} = \frac{ \sum_{i=1}^{n}(x_i-\bar{x})(y_i-\bar{y}) }{%
      \sqrt{\sum_{i=1}^{n}(x_i-\bar{x})^2}\sqrt{\sum_{i=1}^{n}(y_i-\bar{y})^2}}
\label{pcc}
\end{equation}
In Equation~\ref{pcc}, $n$ is the number of data pair, $x_i$ and $y_i$ are data points, $\bar{x}$ and $\bar{y}$ respectively correspond to $\frac{1}{n}\sum_{i=1}^{n}x_i$ and $\frac{1}{n}\sum_{i=1}^{n}y_i$.

The correlation coefficient computed for the data corresponding to the duration as a function of the dex size is equal to \num{0.152}.

With its value close to 0, we can say that the size of the bytecode of an application does not influence the duration of the analysis,
this means that even if an attacker naively introduces libraries or code to bring noise in the analysis, e.g., with dead code, it will not force the analysis to reach the timeout.
Similarly, this type of analysis does not seem to be sensitive about the number of classes in an application as the \emph{PCC} computed for the data corresponding to the duration as a function of the number of classes is equals to \num{0.153}.
The same conclusion can be done as for the dex file size, even with a lot of noise, meaning many classes brought by obfuscation, for example, the analysis still stays efficient.

\begin{figure}
\centering
    \includegraphics[scale=0.33]{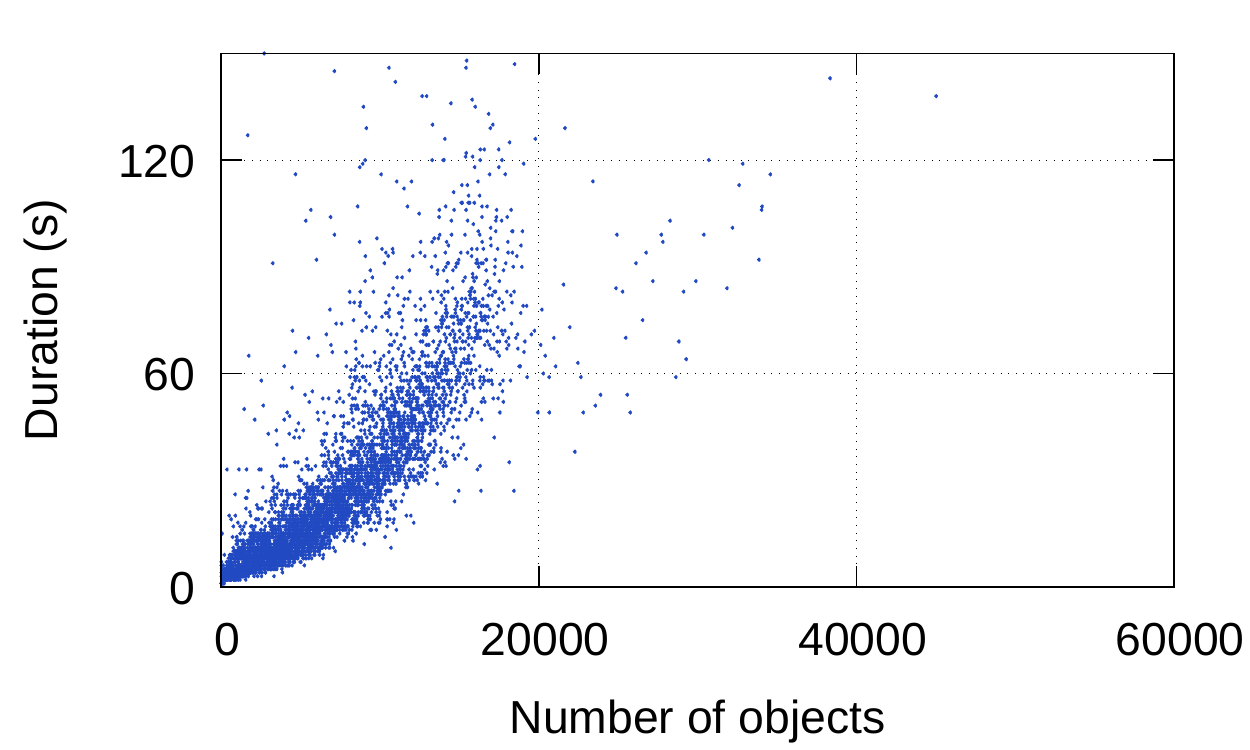}
    \includegraphics[scale=0.33]{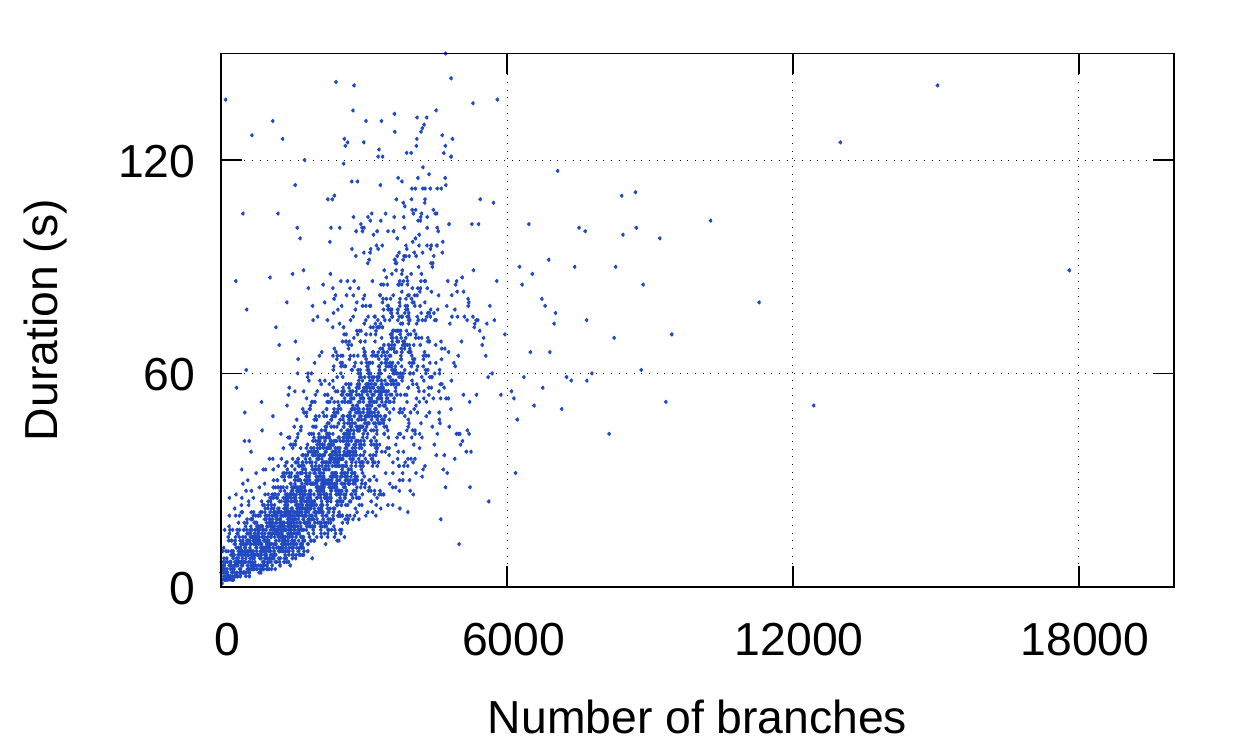}
    \caption{Evolution of the duration of the analysis depending on the number of objects and branches in the applications considered.}
  \label{features_time_exp}
\end{figure}

On the other hand, other features directly influence the duration of the analysis.
In Table~\ref{pcc_scc_to_time} representing the correlation coefficients of Figure~\ref{features_time_exp} we can see that the more objects in an application, the more time it will take to analyze the application.
Indeed, while the \textit{Pearson correlation coefficient} does not indicate any linear correlation (we can intuitively see the exponential correlation in Figure~\ref{features_time_exp}) due to the $PCC$ value of \num{0.359}, the \textit{Spearman Correlation Coefficient} computed based on Equation~\ref{scc} assures us that the relationship between the variables observed can be represented using a monotonic function~\cite{schober2018correlation} due to a coefficient of \num{0.908}.

\begin{equation}
  r_s = \frac{cov(rg_x,rg_y)}{\sigma_{rg_x}, \sigma_{rg_y}}
  \label{scc}
\end{equation}

In Equation \ref{scc} $rg_x$ and $rg_y$ respectively represent the rank variables of $x$ and $y$.
Similarly, $\sigma_{rg_x}$ and $\sigma_{rg_y}$ respectively represent the standard deviations of $rg_x$ and $rg_y$.

Better, as exponential functions can be approximated into linear functions by taking the logarithm of both sides, we can compute a linear correlation coefficient on $(x_i,log(y_i))$ for $i \in \lbrace 0, 1, ..., n \rbrace$ ($n$ being the number of pairs of data) and extrapolate the results for the original data.
We obtain a score of 0.795, which is a strong linear correlation, assuring us that the original data is positively correlated following an exponential function.

The explanation for this exponential correlation is simple: 
to understand and detect logic bombs, 
this approach aims at retrieving the semantic of objects of interest.
This means that the more objects to model, the more statements to take into account while modeling, 
therefore the more time the analysis will take.
This can be problematic for an application with many objects or an application where the developer deliberately introduces useless objects to introduce noise in the analysis.

Additionally, we can see in Table~\ref{pcc_scc_to_time} that the number of reachable branches and the duration of the analysis is, similarly to the number of objects and the duration of the analysis, positively correlated following an exponential function.
Indeed, it introduces new paths, meaning many values to remember depending on the path during the symbolic execution.
Regarding the approach used in this analysis, the most troublesome consequence of having many branches (path minimization is an NP-hard problem) is that it considerably slows down the analysis.

\begin{table}
  \centering
  \begin{tabular}{lrrrr}
    \hline
    & \multicolumn{1}{l}{\textbf{PCC}} & \multicolumn{1}{l}{\textbf{SCC}} & \multicolumn{1}{l}{\textbf{PCC ($x_i,log(y_i)$)}}\\ \hline
    \textbf{\# of objects to time}   & \num{0.359} & \num{0.908} & \num{0.795}\\ \hline
    \textbf{\# of branches to time}  & \num{0.331}& \num{0.839} & \num{0.657} \\ \hline
  \end{tabular}
  \caption{Correlation coefficients of the data of Figure~\ref{features_time_exp} (PCC: Pearson Correlation Coefficient, SCC: Spearman Correlation Coefficient)}
  \label{pcc_scc_to_time}
\end{table}

Furthermore, as the path predicate recovery is not necessary over the entire code of an application, we collected, afterward, on a subset of the large-scale study's applications the average time taken by the predicate recovery step.
It revealed that it is responsible for \num{22.2}\% of the analysis time of an application on average and also responsible, in \num{35.3}\% of the cases, for reaching the timeout.
In contrast, the symbolic execution is responsible for \num{61.7}\% of the analysis time on average, but it has to be performed over the entire application to decide to classify predicates.
It must be taken into account for future work in order to optimize the number of successfully analyzed applications.

For understanding the general scheme in which the logic bombs, even false-positives, are triggered we extracted for each detected trigger the type of component in which the method triggering the logic bomb is located as well as the component where the call stack starts for reaching this method, referred to as starting component.

\begin{figure}
    \centering
    \includegraphics[scale=0.44]{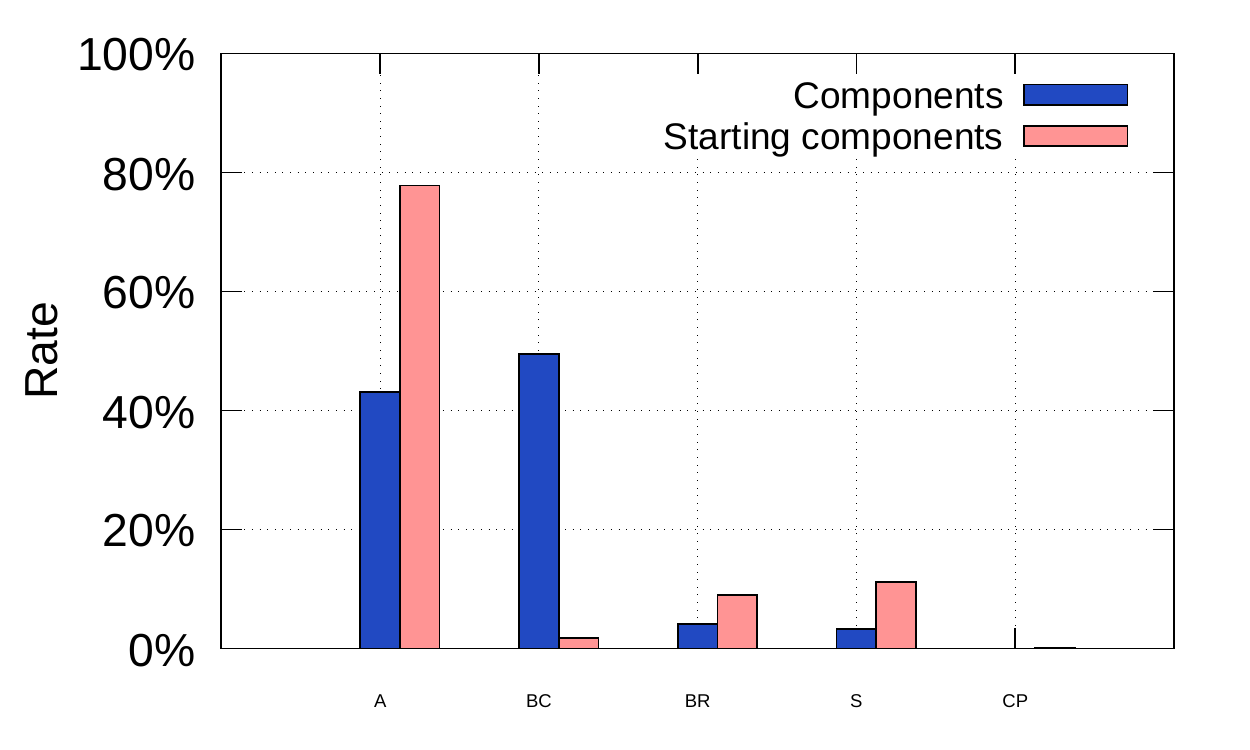}
    \caption{Rate of components (A: Activities, BC: BasicClass, BR: BroadcastReceivers, S: Services, CP: ContentProviders)}
    \label{feature_components_starting_components_call_stack}
\end{figure}

In Figure~\ref{feature_components_starting_components_call_stack} we can see that in a large number of cases, components containing the method triggering the logic bomb are non-Android classes (\num{49.44}\%).

Also, \num{43.1}\% are located in \texttt{Activities}, meaning that the trigger can also be directly embedded in the user code interface.
It makes sense since many applications use time-related triggers for user interfaces (e.g., games).

If we take into account the starting components, it becomes more evident.
In fact, almost 80\% of the starting components are \texttt{Activites}.
Many of these are likely to call a method of another class to trigger the logic bomb.
Another interesting fact in starting components is that, despite the low proportion, the process of triggering often starts in a \texttt{BroadcastReceiver} or a \texttt{Service}.
\texttt{BroadcastReveivers} are, in this case, mostly linked to SMS-related triggers.
Regarding \texttt{Services}, we can assume that this method is likely to be used to monitor the device and trigger code at the right moment.

\begin{figure}
\centering
    \includegraphics[scale=0.40]{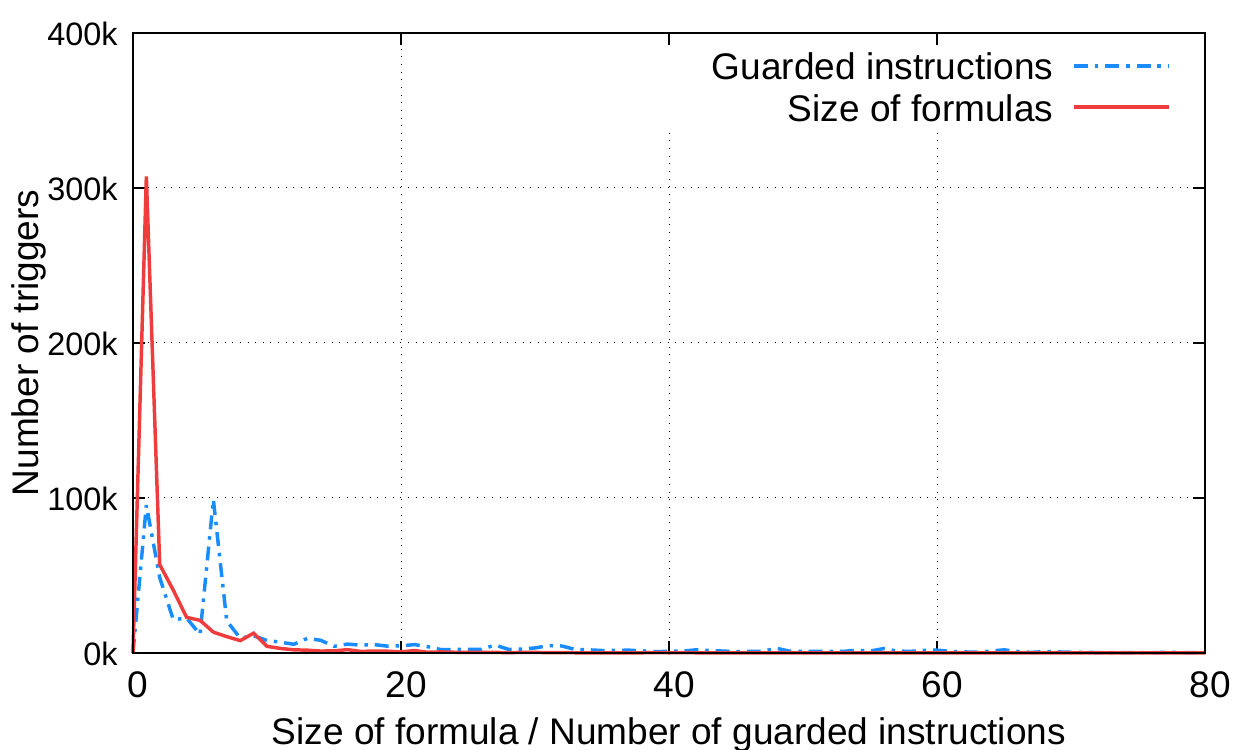}
    \caption{Size of logical formula and count of guarded instructions}
  \label{size_formula}
\end{figure}

We also extracted two other features to understand the form of the check when detected.
We wanted to know if the intra-procedural logic formula extracted during the analysis was complex or not in the general case.
We can see in Figure~\ref{size_formula} that in the majority of the cases, there is only one predicate in the formula, which means that the triggered behavior, in the case of this particular analysis, is generally isolated not part of a multiple branch decision.

Furthermore, the density of instructions dominated by a trigger is interesting to study.
Indeed, we can see that in the majority of cases, the number of guarded instructions by a trigger is lesser than 10 (\textsc{Jimple} instructions).
As the number of instructions is small, we can assume that those instructions represent different calls to other classes' methods to perform a useful action.
This assumption correlates with the fact that in most cases, the component in which is the trigger is a basic class (see Figure~\ref{feature_components_starting_components_call_stack}), that is to say, a non-Android component class.
In fact, in \num{55.29}\% of the cases, one of the instructions is a method call, which confirms our previous data records of Figure~\ref{feature_components_starting_components_call_stack}.

To retrieve the rate of false-positives among the \num{99651} detected applications, we based ourselves on VirusTotal~\cite{total2012virustotal}.
However, the VirusTotal score is challenging to trust for qualifying an application as malware.
That is why we decided to classify these applications by detection rate.
Table~\ref{detection_rate} shows that the rate of false-positives reaches a lower bound of \num{17.2}\% and an upper bound of \num{24.6}\%.

\begin{table}
\centering
\begin{tabular}{cccccccccccc}
\hline
  \multicolumn{1}{c}{\textbf{VT}} & \multicolumn{1}{c}{\textbf{$>$0}} & \multicolumn{1}{c}{\textbf{$>$10}} & \multicolumn{1}{c}{\textbf{$>$20}} & \multicolumn{1}{c}{\textbf{$>$30}} & \multicolumn{1}{c}{\textbf{$>$40}} & \multicolumn{1}{c}{\textbf{$>$50}} 
\\ \hline
  \textbf{Apps} & \num{29829} & \num{15861} & \num{7919} & \num{1237} & 55 & 1
\\ \hline
  \textbf{FP Rate} & \num{17.2}\% & \num{20.6}\% & \num{22.6}\% & \num{24.2}\% & \num{24.5}\% & \num{24.6}\%
\\ \hline
\end{tabular}
  \caption{VirusTotal (VT) detection rate of \tso{} flagged applications (October 2019).}
\label{detection_rate}
\end{table}

\vspace{0.2cm}

\myconclusion{
We applied \tso{} on \num{508122} Android applications with a success rate of 79.9\%.
Our experimentations show that the approach scales on large datasets.
However, it also show that the approach has a high false-positive rate of 17\%
which would require much manual work (which the automated analysis was trying to prevent).
}

\subsection{\textbf{RQ2: What parameters can impact the false positive rate?}}
\label{rq_false_positive_rate}

The conclusion of RQ1 is surprising since we do not reach the false positive rate of the literature (0.3\%).
Thus, in this research question, we identify the main parameters that could significantly impact the false positive rate.
Since we run many analyses, we cannot use the massive dataset of RQ1.
We thus build a new smaller dataset.
In order to build it, we operated as described in the literature.
That is to say, we only considered benign applications from Google Play using the minimum score given by VirusTotal~\cite{total2012virustotal}.
For this, we, again, used the Androzoo dataset~\cite{allix2016androzoo}.
Then, we analyzed the applications to check whether they contained the permission \texttt{android.permi\-ssion.RECEIVE\_SMS}, 
use a location API or a time/date library.

Similarly to the literature, we selected 5803 time-related applications, 
\num{4135} location-related applications, and \num{1400} SMS-related applications.
We ended up with a total of \num{11338} unique benign applications.

\subsubsection{\bf Control Experiment}
\label{sec:controlexpe}

The control experiment, in which we do not change any parameter, 
has been conducted in the same context with the timeout set to 1 hour per app.
Our analysis was able to successfully analyze \num{7297} applications 
out of \num{11338} (i.e., \num{64.4}\%) with an average of 24 seconds per application.
A success means that the analysis for an application did not reach the timeout nor crashed.

The analysis found \num{9535} suspicious triggers, 
\num{4824} applications with a suspicious check, 
\num{3636} applications with suspicious triggered behavior and 
\num{3099} applications after post-filters (see Table~\ref{table_experiments_results} for more information) yielding a false-positive rate of \num{27.3}\%.

\myshortconclusion{
On a dataset with two orders of magnitude smaller than in RQ1, we find that the false positive rate still reaches a high value of more than 27\%.
}

\begin{table*}[]
  \centering
  \begin{tabular}{l|r|r|r|r|r}
	& \textbf{Original results} & \textbf{Timeout 2h} & \textbf{Timeout 3h} & \textbf{Symbolic values filter} & \textbf{Package filter} \\ \hline
	\textbf{\# apps analyzed} & 7297 & 9884 & 9897 & 9880 & 10133 \\ \hline
	\textbf{Mean time of analysis} & 24 & 141.2s & 146.5s & 128.6s & 13.2s \\ \hline
	\textbf{	\# of suspicious triggers} & 5391 & 7724 & 7727 & 1033 & 83 \\ \hline
	\textbf{\# of apps with triggers} & 1701 (23.3\%) & 2373 (20.9\%) & 2376 (21\%) & 381 (3.4\%) & 31 (0.3\%) \\ \hline
  \end{tabular}
  \caption{Experimental results with Timeout variation (cols. 2 and 3), Symbolic Filter (col. 4), Package Filter (col. 5).}
  \label{other_xps_tab}
\end{table*}

\subsubsection{\bf Sensitive Methods Filter}
\label{sensitive_method_filter}
\begin{figure}
\centering
    \includegraphics[scale=0.55]{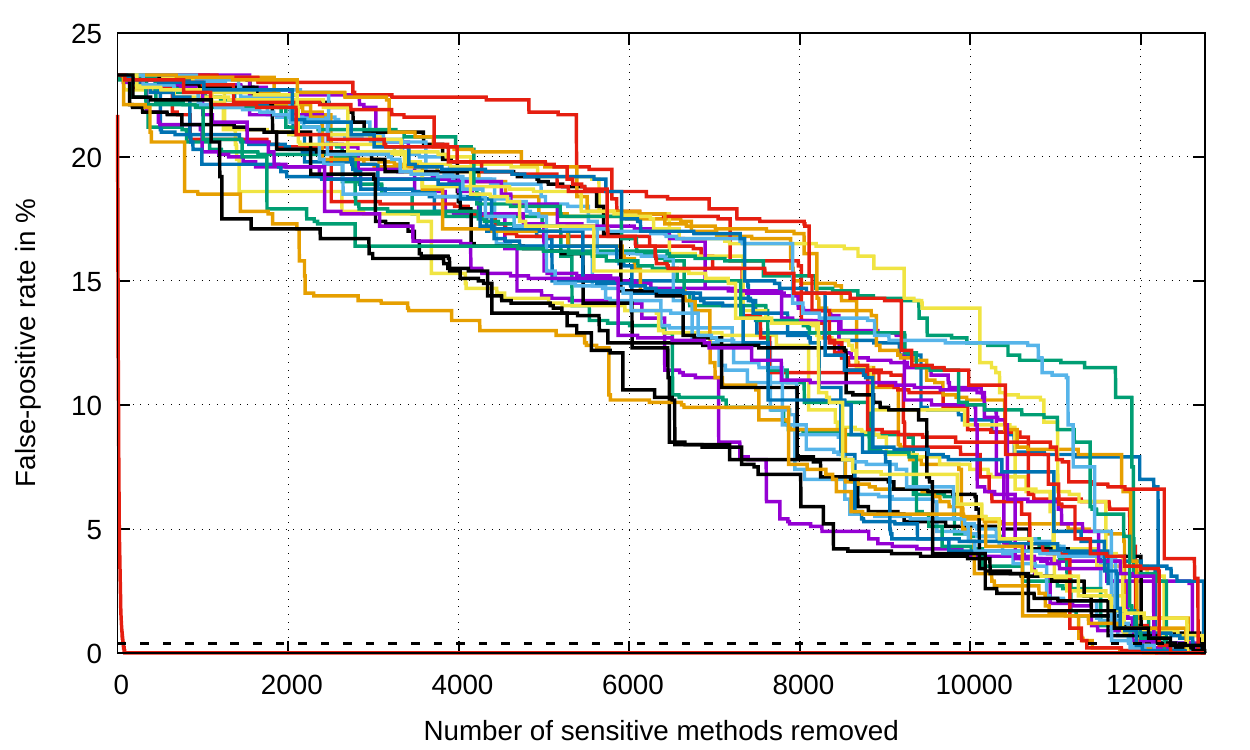}
    \caption{Evolution of the false-positive rate as a function of the number of sensitive methods randomly removed from the list of sensitive methods considered.}
  \label{sensitive_methods_removed-fp}
\end{figure}

In this experiment, we randomly remove methods in the list of sensitive methods, one after the other, to observe the impact on the false positive rate.
We perform this experiment 32 times to see if the results converge.
Figures~\ref{sensitive_methods_removed-fp} shows the results of this experiment.
Each curve represents an experiment.
We see that in order to reach a low false-positive rate (e.g., \num{0.38}\%, represented by the dotted line), 
we have to remove, on average, more than \num{11500} methods ($>$ 90\%) from the list of sensitive methods, which will be missed during the analysis.

We can also see a curve diving fast on the leftmost side of the graph of Figure~\ref{sensitive_methods_removed-fp}.
It represents the same filter for which the most used sensitive methods are removed first.
The sensitives methods are ordered by their occurrence in logic bombs based on the results 
of the control experiment in Section~\ref{sec:controlexpe}.

We observe that to reach a low false-positive rate represented by the dotted line, 
removing the 68 most-used methods is enough.
This means a concentrated number of sensitive methods are used to qualify a trigger as a logic bomb.
Those methods mostly allow one to read device information, write into logs/files, and communicate with the external world.

\begin{table*}[]
  \centering
  \begin{adjustbox}{width=.6\textwidth}
  \begin{tabular}{lccc}
    \hline
    \multicolumn{1}{l}{\textbf{Sensitive Methods}} & \multicolumn{1}{c}{\textbf{Occurences}} & \multicolumn{1}{c}{\textbf{Percentage}} & \multicolumn{1}{c}{\textbf{False-positive rate induced}} \\ \hline
    TextView.setText & 688 & 12.8\% & 1.6\% \\ \hline
    ConnectivityManager.getActiveNetworkInfo & 600 & 11.1\% & 1.2\% \\ \hline
    File.$<$init$>$ & 544 & 10.1\% & 1.2\% \\ \hline
    Log.v & 436 & 8.1\% & 1.6\% \\ \hline
    URL.openConnection & 361 & 6.7\% & 1.3\% \\ \hline
    ContextWrapper.startActivity & 361 & 6.7\% & 1.7\% \\ \hline
    Location.getLatitude & 280 & 5.2\% & 0.9\% \\ \hline
    Log.println & 241 & 4.5\% & 1.0\% \\ \hline
    Activity.finish & 186 & 3.5\% & 0.9\% \\ \hline
    NotificationManager.notify & 163 & 3.0\% & 0.6\% \\ \hline
    File.mkdirs & 118 & 2.2\% & 0.5\% \\ \hline
    Handler.sendEmptyMessageDelayed & 112 & 2.1\% & 1.3\% \\ \hline
    Handler.sendEmptyMessage & 90 & 1.7\% & 0.6\% \\ \hline
    Handler.sendMessage & 85 & 1.6\% & 0.6\% \\ \hline
    TelephonyManager.getDeviceId & 77 & 1.4\% & 0.4\% \\ \hline
  \end{tabular}
  \end{adjustbox}
  \caption{Top 15 sensitive methods considered order by number of occurrence in 
  the defaul experiment of Section~\ref{sec:controlexpe}}
  \label{methods_deleted}
\end{table*}

We provide in Table~\ref{methods_deleted} the list of sensitive methods, 
each present in at least 1\% of logic bombs detected in the control experiment of Section~\ref{sec:controlexpe}.
Those 15 methods represent 80.7\% of the total of potential logic bomb yielded by our tool.
Although some of these methods can be omitted in a definitive list, others like \texttt{TelephonyManager.get\-DeviceId}, which is considered sensitive as it can be leaked and deliver information to the attacker, have to appear in the list of sensitive methods considered.
This method alone corresponds to 0.4\% of the rate of false-positive.
We observe that each method in the list can have a considerable impact on the false-positive rate.

\begin{figure}
\centering
    \includegraphics[scale=0.55]{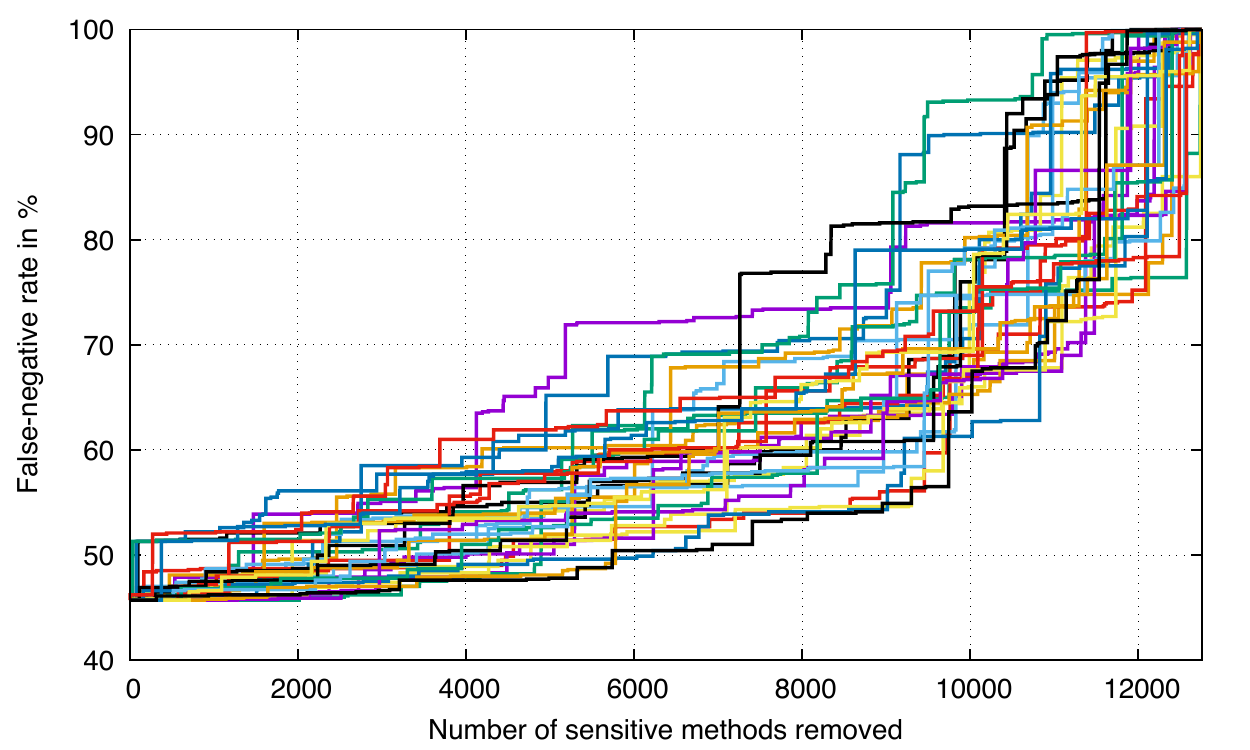}
    \caption{Evolution of the false-negative rate as a function of the number of sensitive methods randomly removed from the list of sensitive methods considered.}
  \label{sensitive_methods_removed-fn}
\end{figure}

From the definition of a false-positive in the literature, we can deduce that a false-negative, 
in this case, it is a malicious application not flagged by the tool.
Therefore, to measure the rate of false-negative in our study, we use a dataset containing only malicious apps.
As before, we randomly remove methods from the list of sensitive methods.
Figure~\ref{sensitive_methods_removed-fn} details our findings.
We can first see that the false-negative rate starts from 45.7\%, then we can see that removing sensitive method increases the rate of false-negative (while decreasing the rate of false-positive, see Figure~\ref{sensitive_methods_removed-fp}).
We have seen previously that removing about \num{11500} sensitive methods 
could be helpful to reach a low false-positive rate close to 0.3\%.
In Figure~\ref{sensitive_methods_removed-fn}, we can see that doing so would 
set the false-negative rate between 70\% and 95\%, which would be unacceptable to detect malicious applications.

\myshortconclusion{
Changing the list of sensitive methods can significantly impact the false-positive rate and the false-negative rate, at least up to two orders of magnitude.
}

\subsubsection{\bf Trigger Filter} 

In this experiment, we modify \tso{} in order not to take into account 
any potential trigger if the values retrieved during the symbolic execution attributed to the test 
were purely symbolic or unknown.
In Table~\ref{other_xps_tab} we can see that the number of suspicious triggers drops to 1033 and the number of applications with suspicious triggers to 381.
This minor change produces results with a factor 5 change regarding the detection rate.
Also, it allows our tool to get a false positive rate close to 3.4\%.
Unfortunately, the analysis misses all logic bombs where triggers are derived from purely symbolic values.

\myshortconclusion{
Using the trigger filter can have a significant impact on the false positive rate.
}

\subsubsection{\bf Package Filter} 

In this experiment, we modify \tso{} in order to only take into account 
methods that are in the same package as the application under analysis.
This filter is stronger than the library filter of Section~\ref{sec:libraryfilter} as it constrains more the analysis.
The results of Table~\ref{other_xps_tab} shows that the number of methods analyzed is significantly reduced.
Indeed, the time taken for the analysis is shallow compared to the other analyses.
Also, the number of triggers yielded by \tso{} reaches \num{83} in \num{31} different apps.
The rate of false-positive is almost equal to the original one.
This heuristic has a significant drawback, and an attacker could easily bypass this filter by changing the package name of classes implementing the triggered behavior.
Unfortunately, the analysis does not take into account all the code outside of the package.
In some applications this accounts for more than 93\% of the code.

\myshortconclusion{
Using the package filter can have a significant impact on the false positive rate.
}

\subsubsection{\bf Library Filter}
\label{sec:libraryfilter}

In this experiment, we filter out well-known libraries in order to remove noise from the results.
For this, we used a list that was made in a study about common libraries~\cite{li2016investigation} used in Android applications.
We manually analyzed 35 of them and confirmed that they contain only false-positives.

After having filtered common libraries from the triggers found beforehand, our results reveal that a scaling approach with this analysis would still not be conceivable concerning the still high number of triggers detected.
Indeed, Table~\ref{results_after_libraries_filter} shows that even with a reduction of \num{43.5}\% of the number of suspicious triggers, there are \num{5391} suspicious triggers.
Also, the number of applications flagged as containing a logic bomb goes from \num{3099} to \num{1701}, a reduction of \num{45.1}\% for the tool but still greater by two orders of magnitude compared to the state-of-the-art.
It means that among \num{11338} unique benign applications there are potentially \num{1701} false-positives (\num{23.3}\%).

Note that, despite being conservative, the false-positive rate calculated during this experiment is obtained by counting the number of benign applications flagged by our tool containing a logic bomb.
This can be explained since a logic bomb necessarily contains malicious code, otherwise, it is triggered behavior.
We acknowledge that even being relatively free from malicious applications, picking applications from \emph{Google Play} is not sufficient to qualify the dataset's applications as benign.
Nevertheless, to stay in line with the literature, we use the same evaluation process.

The majority of detected triggers filtered by the common libraries are time-related triggers.
Out of the \num{4144} suspicious triggers filtered, \num{4065} (\num{98.1}\%) are time-related whereas only 15 (\num{0.36}\%) are SMS-related and 64 (\num{1.54}\%) are location-related.
It shows that common libraries make great use of time-related triggers.
Besides, we have already said that suspicious time-related triggers definition was not narrow enough to detect it compared to SMS-related and location-related.
We can say that even with an efficient library filter, time-related triggers are still commonly used in benign applications.

\myshortconclusion{
The library filter does not have a significant impact on the false positive rate.
}

\begin{table}
\centering
\begin{tabular}{lrr}
\hline
              & \multicolumn{1}{l}{\textbf{Before Lib filter}} & \multicolumn{1}{l}{\textbf{After Lib filter}} \\ \hline
  \textbf{\# Apps w/ LB}   & \num{3099}                       & \num{1701} (\num{-45.1}\%)                  \\ \hline
  \textbf{\# Suspicious LB}  & \num{9535}                       & \num{5391} (\num{-43.5}\%)                  \\ \hline
\textbf{\# LB per App} & 3.1                        & 3.2                       
\\ \hline
  \textbf{\# Time-related} & \num{9099}                        & \num{5034} (\num{-44.6}\%)                       
\\ \hline
  \textbf{\# SMS-related} & 132                        & 117 (\num{-11.3}\%)                       
\\ \hline
  \textbf{\# Location-related} & 304                        & 240 (\num{-21}\%)                       
\\ \hline
\end{tabular}
\caption{Comparison between \tso{}'s results before and after filtering common libraries (LB : Logic Bomb)}
\label{results_after_libraries_filter}
\end{table}

\subsubsection{\bf Different list of sensitive methods} 
\label{different_list}
\begin{table}[]
  \centering
  \begin{tabular}{lc}
    \textbf{\# apps analyzed} & 8285 \\ \hline
    \textbf{Mean time of analysis} & 21.1s \\ \hline
    \textbf{\# of suspicious triggers} & 2855 \\ \hline
    \textbf{\# of apps with triggers} & 956 (11.5\%) \\ \hline
  \end{tabular}
  \caption{Experimental results with different list of sensitive methods.}
  \label{cg_tab_spark}
\end{table}

To build the list of sensitive methods, we reused the results of Pscout~\cite{au2012pscout} and SuSi~\cite{arzt2013susi}, as in the literature.
The constructed list contains \num{12755} methods.
Nevertheless, we have seen that with this list, we obtain a high false-positive rate.
That is why we decided to verify the impact if we were to use another, shorter list of sensitive methods.

We started from the premise that a permission-based method is not necessarily sensitive.
Therefore, we used a list of sink methods from \textsc{Flowdroid}~\cite{arzt2014flowdroid} as they can leak data, which is considered sensitive.
The new list features \num{130} methods.

Nevertheless, the number of triggers flagged by our tool (after re-running the experiment) stays relatively high by reaching 2855.
They are distributed in 956 applications (\num{11.5}\%, see Table~\ref{cg_tab_spark}).
Even with a reduced list of methods considered for the control dependency step, the tool cannot make the difference between malicious and benign behavior.
This shows the need for a more in-depth analysis of the guarded behavior of the triggers.

\myshortconclusion{
Using a reduced list of sensitive methods which are all involved in data leak does not have a significant impact on the false positive rate.
}

\subsubsection{\bf Concept Drift}
Differences in results between \tso{} and existing experiments of the literature could be due to \emph{concept drift}~\cite{jordaney2017transcend}, i.e., the fact that applications used in the experiments of existing papers are older than the ones used in our paper.
We launched experiments on multiple datasets of 10k apps from 2013 
to 2016 and have the following results for the false positive rate: 
2013: 18.5\%, 2014: 15.7\%, 2015: 21.1\% and 2016: 22.6\%.
We observe no significant impact on the results.

\myshortconclusion{
Variation in the application release dates does not have a significant impact on the false positive rate.
}

\subsubsection{\bf Timeout variation} 
Experiments in the literature could have been conducted a couple of years ago.
To simulate the hardware available at the time, we performed the experiments with shorter timeouts.
We launched experiments on multiple datasets of 10k applications and have the following results for the false-positive rate: 30min: 16.1\%, 15min: 15.8\%, and 5min: 15,7\%.
We observe no significant impact on the result.

\myshortconclusion{
Reducing the timeout does not have a significant impact on the false positive rate.
}

\subsubsection{\bf Call-graph construction algorithm} 

The literature might be imprecise and might not always provide all information regarding the implementation of the tool they developed.
Mostly, a crucial part of performing inter-procedural analyses is the call-graph construction algorithm.
Therefore, as we do not always know which call-graph algorithm is used, we renewed our previous experiment by varying the algorithm.
For this, we used the following call-graph construction algorithms: SPARK~\cite{lhotak2003scaling}, CHA~\cite{dean1995optimization}, RTA~\cite{bacon1996fast} and VTA~\cite{sundaresan2000practical}.

Table~\ref{cg_tab} reveals our experimental findings.
First, we can see that none of these algorithms allow us to get a low false-positive rate close to 0.3\%.
It can be deduced that having the correct algorithm will not suffice to have a perfect implementation.
Second, even though the results with Spark algorithm and VTA algorithm are close, we can see that changing the call-graph construction algorithm leads to different results (i.e., false-positive rates of 12.6\% for VTA, 11.4\% for CHA, 13.6\% for RTA, and 11.5\% for SPARK).

Besides, we run the same experiment by increasing the timeout to 2h and 3h.
Table~\ref{other_xps_tab} shows the results of these experiments.
We can see that there is almost no difference between a timeout of 2h and 3h.
However, considering the initial timeout of 1h, the results obtained here are different.
Indeed, the timeout of 1h yielded 1701 applications with triggers, against 2373 and 2376, respectively 2h and 3h.
Likewise, the number of triggers detected increases from 5391 with 1h to 7724 and 7727 with respectively 2h and 3h.

\myshortconclusion{
Changing the call-graph construction algorithm does not have a significant impact on the false positive rate.
}

\vspace{0.2cm}
\myconclusion{
We have experimentally seen that minor changes in the implementation can have an important impact on the results.
Using heuristics allows the approach to get a false-positive rate similar to the literature (0.3\%).
However, this result has a significant impact on the recall, the false-negative rate being raised between 70\% and 95\%.
}

\subsection{\textbf{RQ3: Is it possible to locate the malicious code with logic bomb detection?}}
This is by far the most interesting question for this research area. 
Indeed, detecting malicious code is a difficult problem per se, that is why if this approach could help in this direction, it could be promising.
In fact, \tso{}'s approach is efficient for this purpose for applications taken individually.
Indeed, we manually analyzed 200 apps, and we were able to locate quickly (i.e., in less than 2 minutes on average) the malicious code with the results yielded by \tso{}.
When a true-positive is encountered, we were able to directly inspect the method in which the logic bomb was.
Consequently, we had malicious code at hand.
Note that these manual analyses allowed us to construct \dataset{}, in Section~\ref{trigdb} we give more details.

\vspace{0.2cm}
\myconclusion{
During our numerous manual analyses, we were able to locate/track the malicious code easily. The condition of the logic bomb playing the role of the malicious code entry-point.
}

\subsection{\textbf{RQ4: Do benign and malicious applications use similar
behavior regarding the approach under study and why?}}
\label{rq_similar_behavior}
In this section, we analyze randomly chosen malicious and benign Android applications containing a trigger.

\paragraph{Malicious}
The first malicious application we present is called "LittlePhoto"\footnote{9c92c2279a33de01561ce775c8beee9bbb58895a1f632d19f41ac2b286e12bb2} and allows a person with malicious intents to install third party applications and receive information about the device via HTTP by sending an SMS with \texttt{"\$\$@@\&\&\$\$"} or \texttt{"\$\$@@\&\&@@"} as the content.
It can be viewed as a targeted attack which uses a logic bomb detected as \texttt{\#sms/\#body.equals('\$\$@@\&\&\$\$')}.
This kind of SMS-related logic bomb is usual in remote administration tool (RAT) or SMS-based backdoors.

The second one is called "com.allen.mp"\footnote{54f3c7f4a79184886e8a85a743f31743a0218ae9cc2be2a5e72c6ede33a4e66e} 
and this time relies on a time-related triggered behavior : \texttt{"\#now cmp 14400000L"}.
After decompiling the application and analyzing it (see Appendix~\ref{time_bomb_example} for an example of the code), we found that it checks if there is a ten days period between the current time and a pre-defined value.
If the condition is satisfied, the application retrieves information about the type of operating system, the version of the Android framework, the model of the device, the number of the device, the operator, the type of network, and information about the storage.
Then it sends all this information to a C\&C server: \texttt{"http://search.gongfu-android.com:8511/se\-arch/sayhi.php"}.

\paragraph{Benign}
The first benign application we present is called "Exam Tool"\footnote{c373d79960eadfd34fe56ad67051e2d39536c393513667b9d3c05bcd601dd874}.
It allows students to cheat during an exam without looking at the phone but get the answer from a friend by the number of vibrations the phone would make depending on the received SMS.
\tso{} flagged this application with this logic bomb : \texttt{\#sms/\#body.equals("zebinjo")}, it is clear that it is suspicious according to the definition of a logic bomb.
However, is it malicious? Even if it calls a method categorized as sensitive (on the list of sensitive methods), the answer is no.
Indeed, depending on the SMS received, the phone will vibrate according to a particular protocol defined in the application.
But, 

a hard-coded response code is in the \texttt{onReceive} method of the \texttt{BroadcastReceiver} receiving the SMS.
And when receiving exactly the string "zebinjo" it will trigger the vibration according to the following scheme : "abacdacdcadcdacdacadcacd--ca--c-da-dca-cda-c-ac-a-c-adc-a-c-a-dca-cac-a-dc-ad", the code can be seen in Appendix~\ref{benign_sms_trigger}.

This example shows why \tso{}'s approach cannot be entirely reliable because of the trigger's focus and not the triggered behavior itself.
It implies detecting malicious code, i.e., the problem reduces to detecting malicious code.

Another example of SMS-related triggered behavior is the family of tracker applications allowing users to get information on a device using SMS commands remotely.
The application using this kind of triggered behavior we are going to analyze is "MyCarTracks"\footnote{17bf1d8afc22a40681f833cf442732795f134e3a4b287fd9f99d21db3fa07a81}.
The trigger condition discovered by \tso{} is \texttt{\#sms/\#body.startsWith("GETPOS")} visible in Appendix~\ref{tracking_application}.

The next benign application named "TrackMe"\footnote{a22374fd3a2f6a91ab15d06be29b0a2134a1756d3aa8afe5d7cac8470b418c8d} makes the use of a time-related trigger-based behavior, which is part of the definition of \tso{}'s approach but is clearly legitimate.
The detected logic bomb is \texttt{\#now cmp 15L} which is a comparison between the current date and a numeric value of 15, see Appendinx~\ref{track_me} for the code.
It appears that it is simply a check performed to verify the validity of the trial version of the application.

\paragraph{\dataset{}}
\label{trigdb}
We have previously seen that the large-scale analysis produced a considerable number of false-positives.
Therefore, manually analyzing all the applications flagged by \tso{} is unthinkable.
Nevertheless, we did so on several hundred applications to verify if we could find real logic bombs.

We were able to gather 34 malicious applications containing logic bombs.
We did so by reverse-engineering the applications and starting from the check found by \tso{}.
That way, we had a starting point to track the malicious code.
First, we checked if the application was a known malware in databases like VirusTotal~\cite{total2012virustotal}.
It is not sufficient. Indeed we found many applications respecting this constraint, but the trigger found by the tool was not triggering the malicious code per se.
Second, we tried to understand what was the purpose of the malware (also in databases like~\cite{total2012virustotal}) in order to compare it with the behavior of the code guarded by the trigger found by our tool.
If the behaviors matched, obviously or otherwise, we kept the application for our database.

\begin{table}[]
  \centering
  \begin{tabular}{lccc}
    \hline
    & \multicolumn{1}{c}{\textbf{CHA}} & \multicolumn{1}{c}{\textbf{RTA}} & \multicolumn{1}{c}{\textbf{VTA}} \\ \hline
    \textbf{\# apps analyzed} & 2414 & 3724 & 7605 \\ \hline
    \textbf{Mean time of analysis} & 37.4s & 39.1s & 37.4s \\ \hline
    \textbf{\# of suspicious triggers} & 817 & 1887 & 2925 \\ \hline
    \textbf{\# of apps with triggers} & 275 (11.4\%) & 506 (13.6\%) & 957 (12.6\%) \\ \hline
  \end{tabular}
  \caption{Experimental results with different call-graph construction algorithms and a reduced list of sensitive methods considered. (CHA: Class Hierarchy Analysis, RTA: Rapid Type Analysis, VTA: Variable Type Analyis)}
  \label{cg_tab}
\end{table}

Applications are mainly \textit{Trojans} and \textit{Adwares}.
Each application contains one or more triggers.
These logic bombs are SMS-based and time-based triggers.

We also added 34 benign applications containing trigger behavior similar to malicious triggers.
They mainly come from the Google Play market and are, for the majority, time-related.

For each application, we noted the triggers of interest.
For each trigger, we noted the type and location in the smali code.

We believe this dataset to be useful for the community to encourage future research in finding and understanding logic bombs.
Also, as this dataset is bound to grow, it can be used as a ground-truth to evaluate alternative approaches that aim at detecting logic bombs.

\vspace{0.2cm}
\myconclusion{
	Our manual investigations have shown that benign and malicious applications can use the same code for benign and/or malicious behavior.
	Therefore, in this case, the problem of qualifying malicious code remains.
We make available a list of malicious/benign applications that make a similar usage of trigger-based behavior in \dataset{} in the repository of \tso{} project.
}

\vspace{-0.61cm}
\subsection{\textbf{RQ5: Are \ts{}'s results reproducible?}}
\label{rq_ts_reproducibility}
The experiments have been conducted on two datasets: the first is a dataset of malicious applications, and the second is a dataset of benign applications.
To faithfully reproduce the experiments, we wanted to use the datasets of the experiments in the original paper.
Unfortunately, the list of benign applications has been lost. 
Hence, we created a new dataset which has the same properties as the original dataset.
Concerning malicious applications, 3 out of the 14 considered in their experiments were shared with us.

\subsubsection{Malicious applications}
We have executed \tso{} over the three malicious applications \ts{}'s authors were willing to share to check if the same logic bombs described in their paper could be found.
The first one is called \textit{Holy Colbert}, the second one comes from the \textit{Zitmo} malware family, and the last one is the \textit{RCSAndroid} malware.

First, when executing \tso{} over the application called "Holy Colbert" coming from the so-called MalGenome dataset~\cite{zhou2012dissecting} we effectively find the same time-bomb as they did, but not only, we also discovered an SMS-bomb, see Appendix~\ref{holycolbert} for more details.
The SMS-bomb revealed by our tool is \texttt{\#sms/\#bo\-dy.matches\-("health")} which represents a suspicious narrow check against the body of an incoming SMS.
It is triggered if the time-bomb is satisfied, meaning at a specific date, here the May, 21$^{st}$ 2011.
It triggers the deletion of data through the content resolver.
This implies that our implementation is not entirely identical to the original.
We do not claim that this additional finding makes our implementation more precise because it may introduce more false-positives in other applications.

Finally, regarding the "Zitmo" and "RCSAndroid" malicious applications they provided us, \tso{} was able to extract the same logic bombs as \ts{}.

\subsubsection{Reproducibility of \ts{}'s results}

In this section, we further investigate the discrepancies between \tso{}'s results and \ts{}'s.

In the original paper, they had a total of \num{9582} unique benign applications due to overlap between categories.
We made the list of the \num{11338} applications public in the project repository for reproducibility purposes.

\begin{table}
\centering
\captionsetup{justification=justified}
\begin{tabular}{lrrrr}
\hline
\textbf{Domain} & \multicolumn{1}{l}{\textbf{\# Apps}} & \multicolumn{1}{l}{\textbf{\# w/ SC}} & \multicolumn{1}{l}{\textbf{\# w/ STB}} & \multicolumn{1}{l}{\textbf{\# After PF}} \\ \hline
  \textbf{Time}      & \num{2967} (\num{4950})             & \num{1719} (302)                        & \num{1263} (30)                              & \num{1094} (10)                            \\ \hline
  \textbf{Location}    & \num{3305} (\num{3430})             & \num{2366} (71)                         & \num{1817} (23)                              & \num{1516} (8)                             \\ \hline
  \textbf{SMS}       & \num{1025} (\num{1138})              & 739 (89)                         & 556 (64)                              & 489 (17)                            \\ \hline
  \textbf{Total}       & \num{7297} (\num{9518})              & \num{4824} (462)                         & \num{3636} (117)                              & \num{3099} (35)                            \\ \hline
\end{tabular}
  \caption{Result of our analysis on the \num{11338} benign applications. The values in parenthesis represent \ts{}'s results for the original dataset. (SC: Suspicious Checks, STB: Suspicious Triggered Behavior, PF: Post-Filters).}
\label{table_experiments_results}
\end{table}

First, we observe in Table~\ref{table_experiments_results} a considerable difference between our analysis and \ts{}'s: 
while \tso{} identified \num{3099} suspicious apps, \ts{} identified only 35~\footnote{Unfortunately, \ts{} authors were unable to run their tool on our dataset to compare the results, so we reused the results of their paper on their original dataset to compare our results.}.
While it is true that the two datasets are different, we did not expect to find a two order of magnitude difference between the results.

Second, we extracted, from each application, and each potential suspicious check, different features to understand and verify our results.
Among them, we retrieved the class containing the suspicious check and the method in which it appears, and the sensitive method invoked to flag it.
It appears that only 20 methods in the list of sensitive methods represent 89\% of the sensitive methods considered to flag the suspicious checks.
The list of sensitive methods that we used might explain why we have such a difference in our results.
Nevertheless, according to us, it cannot explain the gap alone because other factors could impact the results.
Consequently, we also analyzed the classes and methods containing suspicious triggers to verify if some distinct pattern might emerge.

We found \num{3165} different combinations of class/method among the \num{9535} suspicious triggers without any combination being overly represented.
We manually analyzed the most used of them (i.e., the first 35) to verify if they were logic bombs.
It seems that they enter the definition of a logic bomb according to the paper in question, but they are not.
In Listing~\ref{connect_to_camera} we can see an example in the \emph{card.io} library, which executes some code if some time has elapsed, then it executes the method called \texttt{android.hardware.Camera.Open()} which is considered suspicious in the list of sensitive methods.
We chose this example to emphasize that most of them are based on time-related triggers.
Better, we found that within the \num{9535} logic bombs found (among the \num{3099} applications flagged), only 3.1\% were location-related and 1.4\% were SMS-related.

We also note that most of the suspicious triggers (\num{43.5}\%) are part of a library used in applications.
Manual analyses revealed that they are not logic bombs, thus introduce noise in the analysis.
\begin{listing}
\begin{minted}{java}
private Camera connectToCamera(int checkInterval, int maxTimeout) {
    long start = System.currentTimeMillis();
    if (this.useCamera) {
        do {try {return Camera.open();}
            catch (RuntimeException e) {...}
        } |\textcolor{red}{while (System.currentTimeMillis() - start < maxTimeout);}|
    } return null;
}
\end{minted}
  \caption{Trigger in io.card.payment.CardScanner class of card.io library (simplified). The \texttt{Camera.open()} method (on the list of sensitive methods) is triggered -not only- under the condition triggered by the \textit{while} instruction.}
\label{connect_to_camera}
\end{listing}
We observe that no filter mechanism is mentioned in \ts{}'s paper.
We contacted the authors, but we could not get the information on whether a filter was used or not in their implementation.
According to our results, to reach such a low rate of false-positives (\num{0.38}\%), at least a library filter has to be used to rule out repeated false-positives.

The reader may have noticed the structure of the previous logic bombs, i.e., nested conditions.
It raises the question of whether this type of structure is often used in trigger-based behavior.
Also, we want to verify if considering nested conditions could have been treated as a single logic bomb by \ts{} developers.
Therefore, it could explain the gap between our results and theirs.
We measured this and found that only \num{16.38}\% of detected triggers have a nested structure.
According to this number, we can conclude that it does not impact the conclusion when comparing \ts{} and \tso{}.

\vspace{0.2cm}
\myconclusion{
We were able to construct a dataset with the same properties as the one used in the original paper.
However, \tso{} yielded a high false-positive rate by detecting \num{3099} potential apps with logic bombs ( $>$ 27\%).

We see the importance of having the original list for reproducing this experiment as it can significantly impact the false-positive rate.
Therefore, with the information that has been provided to us and the description of the approach made in the original paper, we conclude that the approach might be used in a realistic setting to detect logic bombs.
}

\section{Limitations}
\label{limitations}
\textbf{Trigger types.}
Only three trigger types have been modeled, which is not representative of logic bombs, though expanding it to other types would be easy.
Regarding other types of logic bombs, we recently found that a new banking Trojan named "Cerberus" made smart use of the accelerometer sensor for monitoring the device~\cite{cerberus}.
Indeed, it is based on the assumption that a real person would move with its device, hence changing the step counter's values.
Only if this condition is satisfied would the malicious code be triggered.

In a recent analysis, Stone found a malicious application using multiple evading techniques~\cite{thepathtothepayload}.
The malware will first check if the device has a Bluetooth adapter and a name, which is important as emulators use default names.
Then it would check if the device has a sensor and verify the content of \texttt{/proc/cpuinfo} to find both \textit{intel} and \textit{amd} strings.
As most devices use ARM processors, those strings should not appear.
It also checks the appearance of any \emph{Bluestacks} files, which is an emulator solution and other emulation detections.
Finally, this application would deliberately throw an exception and check the content to find any matching string that would show an emulator's existence.

\textbf{Predicate Minimization.}
The next limitation lies in the fact that predicate recovery and predicate minimization are performed systematically, which increases the probability of running into a complicated formula for which the minimization step would never end.
Besides, this step is responsible for \num{22.2}\% of the analysis time and responsible, in \num{35.3}\% of the cases, for reaching the timeout of the analysis.
Unlike the symbolic execution step, which is responsible for the largest part of an application's analysis time (\num{61.7}\%), the path predicate recovery and minimization are not necessary for the entire application.
Indeed, the symbolic execution phase is necessary to decide on the suspiciousness of conditions.
A countermeasure would be to locate interesting checks and then perform the full path predicate recovery and minimization.

\textbf{Maliciousness.}
The most critical weakness of \ts{} approach is the control dependency step, which compares method calls dominated by suspicious triggers with a list of sensitives methods.
This phase requires more attention as it is used to qualify the maliciousness of a condition.
Indeed, a suspicious check can be harmless due to the same usage benign and malicious applications make of trigger-based behavior.
Nevertheless, we recognize the difficulty of this step, given the lack of a formal definition of malware.
Despite having considerable resources, major companies also realize the difficulty to automatically qualify malicious code, e.g., Google still accepts malicious applications in its PlayStore~\cite{googleanubis}.

\textbf{Implementation Errors.}
Even though we reproduced faithfully the approach described in \ts{} paper and reused available and well-tested state-of-the-art code when possible, we are not immune from implementation errors.

\textbf{Implementation Unknowns.}
Given the details in the original paper, we are not able to reproduce the results.
Therefore, we tried to vary parameters and implementation details to get as close as \ts{}'s results.
However, it is challenging to test all the combinations of implementation/parameters to get the original results.

\vspace{-0.4cm}
\section{Related work}
\label{related_work}

In 2008, Brumley \& al.~\cite{brumley2008automatically} developed \textsc{MineSweeper} which is an interesting approach to assist an analyst.
Their solution worked directly at the binary level of an executable application.
Their goal was to uncover trigger-based behavior by constructing conditional paths and input values to execute the application.
The next step was to ask a solver whether the path is feasible or not.
If not, they would not explore this path. On the contrary, they would explore this path and ask the solver to construct -if possible- input values to satisfy the formula.
They would then execute the application with the computed trigger input values to inspect the behavior, and if a malicious behavior were encountered, they would know the conditional path leading to this behavior, thus detecting if a logic bomb exists.
They conducted their experiments on four real-world applications and succeeded in finding trigger-based behavior in less than 30 minutes per application with less than 14 potential logic bombs per application.
Unlike \tso{}, the process is not entirely static nor fully automatic and requires a human to infer the logic bomb.

Four years later, Zheng \& al.~\cite{zheng2012smartdroid} focused on finding a user interface-based trigger that could be used to hide malicious code to traditional analysis in Android applications. 
They construct what they call the FCG (Function Call Graph) to retrieve call paths to sensitive Android APIs.
The next step is constructing what they call the ACG (Activity Call Graph) to have the relationship between the application activities, i.e., how to go from one activity to another via user interface methods.
Having those details, they run the application by triggering user interface elements to go to the sensitive activity, which calls a sensitive API and monitors the behavior to check if anything suspicious happens.
That way, they can deduce if the user interface triggering process is used to trigger malicious code.
Their approach is not generic and focuses on one single type of logic bomb.
Also, we note the use of dynamic analysis of their approach again.

Pan \& al.~\cite{pan2017dark} presented a new machine-learning based technique to detect \textit{Hidden Sensitive Operations} in Android apps.
They do not specifically focus on malicious behavior contrary to \ts{}'s approach, which targets malicious activities.
Their approach is composed of a pre-processing part where lightweight data-flow and control-flow analysis are performed to extract a condition-path graph.
This latter is then used to extract features that will feed the SVM classification.
Doing so, \textsc{HSOMiner} performs a precision of 98.4\% (based on 125 randomly chosen apps from a set of \num{63372}) and a recall of 94\%.
Though the approach is interesting (SVM is resistant to overfitting), it does not fit the goal of detecting logic bombs hidden in Android apps.

In 2017, Papp \& al. tried to work directly at the source code level to detect trigger-based behavior in legitimate applications~\cite{papp2017towards}.
Their goal is not to work on a malicious application.
They want to emphasize triggered behavior or backdoor behavior in legitimate open-source applications.
For this purpose, they use \textsc{KLEE}~\cite{cadar2008klee} to perform mixed concrete and symbolic execution (also called concolic execution).
However, first, they have to instrument the considered application to add specific library calls to use these existing tools.
Then they execute the concolic execution and based on the result, and they generate different test cases.
Out of these test cases, some can be highlighted by their program to be verified by an analyst to check whether it is a trigger-based behavior or not.
Though it is promising for a semi-automatic detection tool, their approach can take a large amount of time to generate test cases and lead to a large number of false-negatives.

We now present a more advanced and promising method developed by Bello and Pistoia~\cite{bello2018ares}.
Their approach aims at exposing evading techniques that sophisticated malware use.
Nevertheless, they do not stop at the detection, as we will see.
Their work is divided into three parts, the first one being the detection of \emph{evasion point candidate} by using information flow analysis.
They use the notion of source and sink, that is to say, detecting information going from a source and going to a sink.
Once detected, they step into stage two, it is simply the instrumentation of the Java bytecode to force the untaken branch during the following analysis.
The last stage is precisely the execution of the instrumented application in a controlled environment to monitor the malicious code's behavior.
According to the authors, their approach is unsound but a stepping stone toward detecting new malware behavior.
The first stage of their work is related to our work, except that they follow the flow of fingerprinting methods to branches.

Logic bombs can be used for venerable purposes, indeed in a recent study, Zeng \& al.~\cite{zeng2019resilient} presented an approach to detect repackaging using triggers.
Indeed, their approach consists of instrumenting a legitimate app that could be repackaged by an attacker and adding an instruction to make the app "repackage-proof".
They introduce cryptographically obfuscated trigger conditions that, when triggered, can detect if the program has undergone a repackaging process.
The term "logic bomb" specifically means triggering malicious code under specific circumstances. However, the code triggered is not malicious but preventive. Therefore, although they use the same mechanism as malware developers, the authors should talk about \textit{Hidden Preventive Code}.
Nonetheless, their approach is resilient since we assume the condition has been detected, which is already a challenging problem. One cannot resolve the condition due to its cryptographic properties (using hash functions). Therefore the guarded code cannot be decrypted and executed.
Inserting checks in legitimate apps is promising for protecting Android apps from malware developers and has well been studied in the literature~\cite{luo2016repackage}.

\section{Conclusion}
\label{conclusion}

In this paper, we have implemented \tso{}, the first open-source version of 
the state-of-the-art approach for detecting logic bombs in Android applications.
We first conducted a large-scale analysis over a set of more than \num{500000} Android applications and observed that the approach scales.

However, the approach is not appropriate for {automatically} detecting logic bombs 
because the false-positives rate of 17\% is too high.
We conducted multiple experiments on the approach's parameters to understand the impact on the false positive rate and identify that a low false-positive rate could be reached but at the expense, for instance, of missing a large number of sensitive methods.
That is to say, the approach with a low false-positive rate misses a large number of logic bombs.
We experimentally show that \tso{}'s approach might not be usable in a realistic setting to detect logic bombs with the original paper's information. 

Moreover, we have seen that \tso{}'s approach is not sufficient to detect logic bomb because benign and malicious apps can use the same code for benign and/or malicious behavior. This is a direct consequence of the lack of a formal definition of what a malware is.
We empirically show that using \tso{}'s approach, \emph{trigger analysis} is not sufficient to detect logic bombs.
Indeed, dissociating the trigger condition and the guarded behavior produces false-positives.
Thus, an analyst is necessary to verify the behavior.
Nevertheless, when manually inspecting malicious applications containing logic bombs, provided by \tso{}, we were quickly able to verify if the triggered code is malicious.
We did not have to search through all the code, which saved us a lot of time.
Hence, \tso{}'s approach seems promising to locate the malicious code using logic bombs 
in applications being reverse engineered.

We created \dataset{}, a database of 68 applications containing localized trigger behavior.
We manually analyzed and classified each logic bomb.
We make it publicly available for future research as a ground-truth promised to evolve.

\vspace{.5cm}

In a nutshell, our work shows that, even though \tso{}'s approach is interesting, it is not suitable for automatically detecting logic bombs.
Indeed, the control dependency step is not sufficiently representative of malicious behavior.
Our results contradict state-of-the-art by two orders of magnitude regarding the rate of false-positives.
We have identified several parameters, such as the list of sensitive methods, which could have a two-order magnitude impact on the false positive rate. These parameters should be detailed in every paper tackling the challenging task of detecting logic bombs in Android applications.  
Our hope is that future publications do not omit this information to makes their experiments reproducible.

\section{Future work}
\label{future_work}
We have seen that if an application uses trigger-based behavior combined with malicious code, they are inseparable, sophisticated techniques should be used to qualify the behavior as malicious to flag it as a logic bomb.
\ts{} checks if a sensitive method is called, but an in-depth analysis of the guarded behavior is to be preferred to understand the triggered behavior.
Because a sophisticated analysis of the guarded behavior reduces to detecting malicious code, the difficulty remains due to the lack of a formal definition of what malware is.
We aim at going further by linking the guarded code with its condition.
The trigger would somehow be the entry point for the code. Hence we would be able to locate the malicious code, or at least be able to track it.
We also want to verify if the guarded code is enough to execute a behavioral study and qualify the application's maliciousness.

We do not just want to focus on the internal context of the application.
Indeed, the external context, i.e., the application description or the application reviews, can be processed by applying NLP algorithms (\cite{jivani2011comparative},~\cite{blei2003latent}).
Existing techniques such as~\cite{gorla2014checking} and~\cite{qu2014autocog} already use the application descriptions to compare it with the permissions or the application's behavior.

Finally, we aim to build a classification of types of triggers to understand their use in malicious applications.
This would help build a detector that could take into account those classes and study their use in the wild.

\section{Acknowledgments}
\label{acknowledgments}
This work was supported by the Luxembourg National Research Fund (FNR)
(12696663 and 14596679). 
This work was partially supported by the Wallenberg AI, Autonomous Systems and 
Software Program (WASP) funded by the Knut and Alice Wallenberg Foundation.
The authors would like to thank Dr. Francine Herrmann (University of Lorraine,
Dr. Yann Lanuel (University of Lorraine) and Dr. Imen Sayar (University
of Luxembourg) who have provided insightful comments for writing this paper.
Also, the authors want to thank Prof. Loïc Colson (University of Lorraine)
and Timothé Riom (University of Luxembourg) for their
qualitative reflection on technical aspects of this paper.

\bibliographystyle{IEEEtran}
\bibliography{bib}

\begin{IEEEbiography}[{
\includegraphics[width=1in,height=1.2in,clip,keepaspectratio]{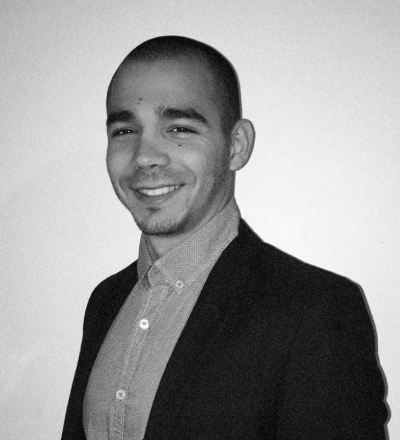}}]
{Jordan Samhi} received his Master's degree in Computer and Information Systems Security 
from the University of Lorraine (France), in 2019. He is, in 2021, a Doctoral Researcher 
at the University of Luxembourg. His research interests are in the security aspects of 
software engineering, with a particular focus on malware and vulnerability detection.
\end{IEEEbiography}

\begin{IEEEbiography}[{
\includegraphics[width=1in,height=1.2in,clip,keepaspectratio]{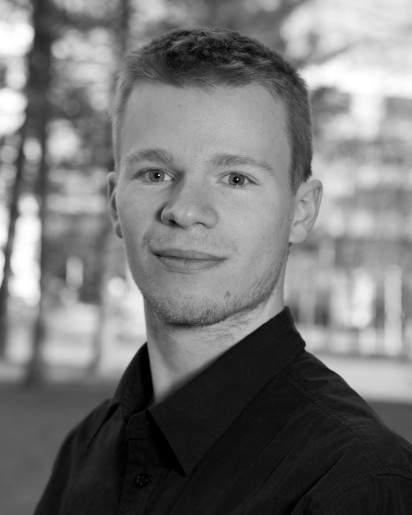}}]
{Alexandre Bartel} is a full professor in the computing science department at Umeå University. 
His current research interests include and combine software testing, Android security and vulnerability analysis.
\end{IEEEbiography}

\onecolumn

\appendices

\section{Implementation of \tso{}}
\label{implementation}
\begin{multicols}{2}
Any implementation is subject to erroneous code.
In order to reduce the number of errors in \tso{}, we rely on well-tested state-of-the-art publicly available Java frameworks. 
\ts{}, on the other hand, is written from scratch in C++ which increases the risks of introducing numerous implementation errors~\cite{phipps1999comparing}.

\textsc{TriggerScope} has originally been developed in C++ with full management of the transformation of the Dalvik bytecode in a custom intermediate representation on which is performed the control flow analysis and the analysis.
The modeling of Android applications has well been explored in state-of-the-art, that is why we did not re-implement those aspects.

\textsc{TSOpen} consists of more than 5K Java SLOC (\num{18.6}K for \ts{}) which, provides better results in term of execution time than the C++ version of \textsc{TriggerScope}.
This is probably due to the choice, on their side, to implement the control flow analysis from scratch.
Also, we parallelized, using multi-threading when it was possible, e.g., the symbolic execution and the path predicate recovery.

Besides using \textsc{Flowdroid} for modeling Android applications, our implementation is built on top of \textsc{Soot}~\cite{vallee2010soot} which is the state-of-the-art solution regarding static analysis over Java and Android~\cite{bartel2012dexpler} programs, initially described in 1999 and since used by researchers around the world.

    Note that we do not possess the information about the call-graph construction algorithm used by the authors of \ts{}.
Therefore, even though we try to faithfully implement the tool with the details of the paper, we cannot have a 100\% similar tool.
\tso{} relies on Flowdroid to construct the call-graph used for the inter-procedural analysis.
Flowdroid, in turn, relies by default on the Spark call-graph analysis framework~\cite{lhotak2003scaling}.
We evaluate \tso{} using different call-graph construction algorithms as discussed in section \ref{evaluation}.

\vspace{-0.5em}

\subsection{Symbolic execution}
Similarly to \textsc{TriggerScope}, our symbolic execution engine models numeric, string, time, SMS, and location-related objects.
The \textsc{Jimple} intermediate representation is, again, convenient for recognizing objects and operation performed over those objects thanks to its flat representation of operations and its explicit typed variables.
Furthermore, as our approach is built over Soot which allows optimizing the analyzed code, numeric and string constants are easily propagated which facilitates the predicate classification.

Modeling string objects is important for the detection of suspicious checks against any string value/field of an object, e.g., the body of an incoming SMS.
For this purpose, our analysis models as faithfully as possible string values propagated along the graph.
When dealing directly with concrete values, the analysis recognizes the operations performed and executes it directly, e.g., \texttt{append(), format(), subString()}, otherwise if it is a symbolic value it records the operation.

Regarding time-related objects, \tso{} keeps track of a list of time-related classes (e.g., \texttt{Calendar, Date, LocalDateTime, SimpleDateFormat, etc.}) and annotates each one with the tag \texttt{\#now} when it recognizes that it has been instantiated with the current date in the program.
Also, \tso{} keeps track of related methods that narrow the circumstances of the potential check performed on such values like \texttt{Date.getHour()} would be annotated with \texttt{\#now/\#hour} which eases the future classification.

Equivalently, \tso{} records every location-related objects like \texttt{android.location.Location} and annotates it with the \texttt{\#here} tag when it is instantiated to represent the current location.
Fields of those objects can be accessed to represent more precise values like the longitude or the latitude. \tso{} annotates those values with respectively \texttt{\#here/\#longitude} and \texttt{\#here/\#latitude}.

Furthermore, the analysis does the same approach for the \texttt{SmsMessage} object, i.e. it annotates it with the \texttt{\#sms} tag and records, as strings, values retrieve from the received SMS.
In this case, it keeps track of the use of the \texttt{getMessageBody(), getDisplayMessageBody(), getOriginatingAddress()} and \texttt{get\-Dis\-play\-Ori\-gi\-natingAddress()} methods 
and annotates them with \texttt{\#sms/\#body} or \texttt{\#sms/\#sender}.

\tso{} also models boolean values to keep track of methods returning a boolean value used in conditions.

For the context of this analysis, \tso{} annotates methods like \texttt{Date.after(), Date.before(), String.contains(), String.startsWith()}, etc.

\subsection{Predicate recovery}
This step aims at getting rid of false dependencies while retrieving instructions dominated by the trigger condition which will be used for the control dependency.
For this purpose, the analysis first begins with a forward intra-procedural analysis and it extracts simple predicates from each check and annotates the edge corresponding with the predicate.
An edge annotated with a predicate \textsl{p} means that the target of the edge must be executed if and only if \textsl{p} is satisfied. 

The next step is the real predicate recovery as it, for each node retrieves the list of predicates to be satisfied to reach this node.
To this end, \tso{} performs a backward intra-procedural analysis and combines, recursively, the previous simple predicates together.
More precisely, a boolean formula is built following two rules: 1) if the node in question has one predecessor, it is combined using a logical AND; 2) if the node in question has more than one predecessor, it is combined using a logical OR between every possible path predicate.

Finally, the last step, without which the analysis would lead to a significant increase of false positives due to false dependencies, is the minimization of boolean formulas.

\subsection{Predicate classification}
Hitherto \tso{} has modeled interesting objects related to the purpose of this analysis and has propagated their values.
It has also removed false dependencies, it now needs to decide whether a check is considered as suspicious.

\noindent\paragraph{Time-related objects} 
\tso{} verifies that 
(1) one of the operands is the result of the invocation of a comparison between two date/time objects such as \texttt{after()} or \texttt{before()} and
(2) one of them representing the current date and the other being built with a constant value.
Similarly, it verifies that, in the check, some primitive numeric types representing the current date/time are compared with a constant.
In these cases, it flags the check as suspicious.

\noindent\paragraph{Location-related objects}
Our tool verifies if one of the operands is a value derived from an object related to the current location and if the other operand represents a constant value.
Also, it checks if one of the operands is the result of the invocation of a method such as \texttt{distanceBetween()} to check if the device is in a specific area.
In these cases, it also flags the check as suspicious.

\noindent\paragraph{SMS-related objects}
\tso{} verifies if one of the operands represents a value from the body of an SMS or the sender of an incoming SMS.
Then, if it is the case, it verifies if these values, which are strings, are matched against specific patterns or constants through the invocation of methods such as \texttt{startsWith(), endsWith(), contains(), matches(),} etc.
These checks are also flagged as suspicious because they encode tight conditions which could be used to exfiltrate data surreptitiously.

Furthermore, to set aside obvious non-suspicious checks like null check reference or the comparison of a number with "-1" (e.g., is the size of the body of an SMS greater than -1 ?), we apply a post-filter step as described in \textsc{TriggerScope} paper.

\subsection{Control dependency}
Finally, we have a reduced list of suspicious checks, we also have the list of instructions that are guarded by each check, the analysis can now determine if any of these checks contain an invocation to a sensitive Android API.

For this purpose, \tso{} iterates over guarded instructions of previously flagged suspicious trigger conditions and checks if they contain an invocation to a method, if it is the case, it verifies if this method appears in the list of sensitive methods considered in the paper of \textsc{TriggerScope}.
The list is not public and not shared but the researchers wrote that they used the result of PScout~\cite{au2012pscout} and SuSi~\cite{arzt2013susi}.
We retrieved these lists as well and used them to classify a method as sensitive.
If a match is found, the check is flagged as a potential logic bomb.

Additionally, this approach is inter-procedural, meaning that the analysis will propagate to analyze the content of any method invocation to check if, in the call stack, a match can be found.
Also, some malware does not invoke a sensitive method directly.
Indeed, the logic bomb could be used to turn a switch (e.g., a boolean) and a check could be performed on this switch elsewhere in the code.
That is why the analysis follows these updated fields between methods and checks whether they are guarded by a check which would invoke a sensitive method.
\end{multicols}

\section{Holy Colbert decompiled}
\label{holycolbert}
\begin{listing}[h]
\begin{minted}{java}
StringBuilder date = new StringBuilder(new SimpleDateFormat("MMddyyyy").format(new Date()));
SmsMessage sms = SmsMessage.createFromPdu((byte[]) ((Object[]) intent.getExtras().get("pdus"))[0]);

if (date.toString().matches("05212011")) {
	int nextInt = new Random().nextInt(4) + 1;
	if (sms.getMessageBody().matches("health")) {
		deleteContent();
	}
	SmsManager.getDefault().sendTextMessage(sms.getOriginatingAddress(), null, new String[]{"Cannot talk right now, the world is about to end", "Jebus is way over due for a come back", "Its the Raptures,praise Jebus", "..."}[nextInt], null, null);
}
\end{minted}
\caption{Holy Colbert application decompiled (simplified)}
\end{listing}

\section{Time-related trigger in tracking application}
\label{track_me}
\begin{listing}[h]
\begin{minted}{java}
if ((System.currentTimeMillis() - Config.Review) / 86400000 < 15) {
	builder = new Builder(MainActivity.this);
	builder.setTitle("Trial expired");
	builder.setPositiveButton("Yes", new DialogInterface.OnClickListener() {
		public void onClick(DialogInterface dialogInterface, int i) {
			Intent intent = new Intent(MainActivity.this, BillingActivity.class);
			MainActivity.this.startActivity(intent);
		}
	});
}
\end{minted}
\caption{Track Me application decompiled (simplified)}
\end{listing}

\section{Android framework model constructed by \textsc{Flowdroid}}
\label{android_model_flowdroid}
\begin{figure}[h]
\begin{tikzpicture}[scale=0.8]

    \tikzstyle{bgo} = [fill = white, fill opacity = 0.8, text opacity = 1]
    \tikzstyle{zoom}= [red, loosely dotted]

    \node[draw,scale=0.7] (onCreate) at (-1,0.65) {onCreate()};
    \node[draw,scale=0.7] (onStart) at (-1,-0.15) {onStart()};
    \node[draw,scale=0.7] (onResume) at (-1,-0.95) {onResume()};
    \node[draw,scale=0.7,diamond,fill=gray!40] (p1) at (-1,-1.85) {$p_1$};
    \node[draw,scale=0.7,diamond,fill=gray!40] (p2) at (-1,-2.86) {$p_2$};
    \node[draw,scale=0.7] (onPause) at (-1,-3.75) {onPause()};
    \node[draw,scale=0.7,diamond,fill=gray!40] (p3) at (-1,-4.65) {$p_3$};
    \node[draw,scale=0.7] (onStop) at (-1,-5.55) {onStop()};
    \node[draw,scale=0.7,diamond,fill=gray!40] (p4) at (-1,-6.45) {$p_4$};
    \node[draw,scale=0.7, diamond,fill=gray!40] (p5) at (-1,-7.45) {$p_5$};
    \node[draw,scale=0.7] (onDestroy) at (-1,-8.35) {onDestroy()};
    \node[draw,scale=0.7] (onRestart) at (1,-7.45) {onRestart()};
    \draw[fill=black] (-1,1.30) circle (0.1);
    \draw[fill=black] (-1,-9) circle (0.1);
    \draw (-1,2) node[] {Component's dummy main method};

    \node[draw,scale=0.8,rounded corners,fill=green!20] (launched) at (7,0.25) {Launched};
    \node[draw,scale=0.8, bgo] (create) at (7,-0.75) {onCreate()};
    \node[draw,scale=0.8, bgo] (start) at (7,-1.75) {onStart()};
    \node[draw,scale=0.8, bgo] (resume) at (7,-2.75) {onResume()};
    \node[draw,scale=0.8,rounded corners,fill=green!20] (running) at (7,-3.75) {Running};
    \node[draw,scale=0.8, bgo] (pause) at (7,-4.75) {onPause()};
    \node[draw,scale=0.8, bgo] (stop) at (7,-5.75) {onStop()};
    \node[draw,scale=0.8, bgo] (destroy) at (7,-6.75) {onDestroy()};
    \node[draw,rounded corners,fill=green!20,scale=0.8] (shutdown) at (7,-7.75) {Shutdown};
    \node[draw,rounded corners,fill=green!20,scale=0.8] (killed) at (4.5,-3.75) {Process killed};
    \node[draw,scale=0.8, bgo] (restart) at (9.5,-1.75) {onRestart()};
    \draw (7,2) node[] {Component life-cycle};

    \draw[fill=black] (-10,0) circle (0.1);
    \node[draw,diamond,fill=gray!40] (p10) at (-10,-1.5) {$p_a$};
    \node[draw,diamond,fill=gray!40] (p11) at (-10,-3) {$p_b$};
    \node[draw,diamond,fill=gray!40] (p12) at (-10,-4.5) {$p_c$};
    \node[draw,diamond,fill=gray!40] (p13) at (-10,-6) {$p_d$};
    \draw[fill=black] (-10,-7.5) circle (0.1);
    \node[draw, bgo] (c1) at (-7.5, -2.25) {Component 1};
    \node[draw, bgo] (cx) at (-7.5, -3.75) {Component X};
    \node[draw, bgo] (cn) at (-7.5, -5.25) {Component N};
    \draw (-10,2) node[] {Application's};
    \draw (-10,1.5) node[] {dummy main method};

    \draw[->,>=latex] (launched) -- (create);
    \draw[->,>=latex] (create) -- (start);
    \draw[->,>=latex] (start) -- (resume);
    \draw[->,>=latex] (resume) -- (running);
    \draw[->,>=latex] (running) -- (pause);
    \draw[->,>=latex] (pause) -- (stop);
    \draw[->,>=latex] (stop) -- (destroy);
    \draw[->,>=latex] (destroy) -- (shutdown);
    \draw[->,>=latex] (pause) -| (killed);
    \draw[->,>=latex] (stop) -| (killed);
    \draw[->,>=latex] (killed) |- (create);
    \draw[->,>=latex] (pause) -- (9, -4.75) |- (resume);
    \draw[->,>=latex] (stop) -| (restart);
    \draw[->,>=latex] (restart) -- (start);

    \draw[->,>=latex] (-1,1.3) -- (onCreate.north);
    \draw[->,>=latex] (onCreate) -- (onStart);
    \draw[->,>=latex] (onStart) -- (onResume);
    \draw[->,>=latex] (onResume) -- (p1);
    \draw[->,>=latex] (p1) -- (p2);
    \draw[->,>=latex] (p2) -- (onPause);
    \draw[->,>=latex] (onPause) -- (p3);
    \draw[->,>=latex] (p3) -- (onStop);
    \draw[->,>=latex] (onStop) -- (p4);
    \draw[->,>=latex] (p4) -- (p5);
    \draw[->,>=latex] (p5) -- (onDestroy);
    \draw[->,>=latex] (p5) -- (onRestart);
    \draw[->,>=latex] (p1.west) -| (-3,-0.95) -- (onResume.west);
    \draw[->,>=latex] (p3.east) -- (1,-4.65) |- (onResume.east);
    \draw[->,>=latex] (p4.west) -- (-4,-6.45) |- (onCreate.west);
    \draw[->,>=latex] (onRestart.east) -- (2,-7.45) |- (onStart.east);
    \draw[->,>=latex] (onDestroy) -- (-1,-8.9);

    \draw[->,>=latex] (-10,0) -- (p10);
    \draw[->,>=latex] (p10) -- (p11);
    \draw[->,>=latex,dotted] (p11) -- (p12);
    \draw[->,>=latex] (p12) -- (p13);
    \draw[->,>=latex] (p13) -- (-10, -7.4);
    \draw[->,>=latex] (p10.east) -- (c1.north west);
    \draw[->,>=latex] (c1.south west) -- (p11.east);
    \draw[->,>=latex,dotted] (p11.east) -- (cx.north west);
    \draw[->,>=latex,dotted] (cx.south west) -- (p12.east);
    \draw[->,>=latex] (p12.east) -- (cn.north west);
    \draw[->,>=latex] (cn.south west) -- (p13.east);
    \draw[->,>=latex] (p13.west) -- (-11,-6) |- (p10.west);
    
    \node[fit=(cx), draw, red, dashed] (a1) {};
    \node[fit={(-4.5,1.5) (2.5,1.5) (2.5,-9.2) (-4.5,-9.2)}, draw, red, dashed] (a2) {};

    \begin{pgfonlayer}{bg} 
    \draw[zoom] (a1.north west) -- (a2.north west);
    \draw[zoom] (a1.north east) -- (a2.north east);
    \draw[zoom] (a1.south west) -- (a2.south west);
    \draw[zoom] (a1.south east) -- (a2.south east);
    \end{pgfonlayer}
    
\end{tikzpicture}
\caption{The left diagram represents the dummy main method of the entire application constructed by \textsc{Flowdroid} with each opaque predicate $p_a$, ..., $p_d$ in gray.
Opaque predicates will not be evaluated during analysis, hence both branches would be considered equally.
Each component life-cycle is modeled after the corresponding life-cycle. 
For instance, if Component X is an Activity component, FlowDroid models it according to the Activity life-cycle presented on the right hand side of the figure.
As for the dummy main, FlowDroid's concrete implementation of the component life-cycle (middle) contains opaque predicates.}
\end{figure}
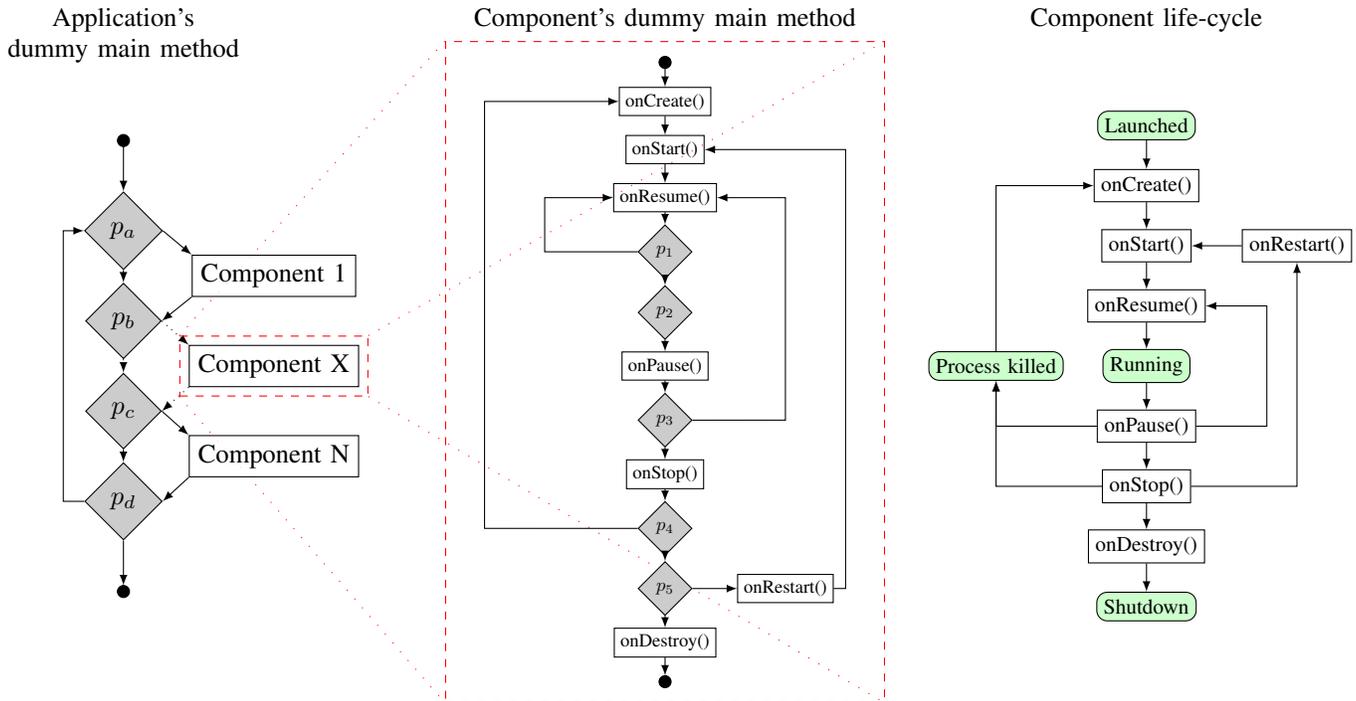

\section{Time-bomb example}
\label{time_bomb_example}
\begin{listing}[h]
\begin{minted}{java}
public void onCreate() {
    long j = sharedPreferences.getLong("start", 0);
    long currentTimeMillis = System.currentTimeMillis();
    if (currentTimeMillis - j < 14400000) {stopSelf();}
    else {
        if (Utils.isConnected(this)) {doSearchReport();}
        getPermission();
        provideService();
    }
}
\end{minted}
\caption{Time-bomb in com.allen.mp application (simplified)}
\end{listing}

\section{Benign SMS trigger}
\label{benign_sms_trigger}
\begin{listing}[h]
\begin{minted}{java}
public void onReceive(Context context, Intent intent) {
    SmsMessage[] smsMessageArr = (SmsMessage[]) null;
    String str = getMessageBody(smsMessageArr);
    String code = "abacdacdcadcdacdacadcacd--ca--c-da-dca-cda-c-ac-a-c-adc-a-c-a-dca-cac-a-dc-ad";
    if (str.equals("zebinjo")) {
        CrnacVibrator crnacVibrator = new CrnacVibrator();
        crnacVibrator.odgovor(code);
        abortBroadcast();
    }
}
\end{minted}
\caption{Exam tool decompiled (simplified)}
\end{listing}

\section{Tracking application}
\label{tracking_application}
\begin{listing*}[h]
\begin{minted}{java}
if (intent.getAction() != null) {
    Bundle extras = intent.getExtras();
    if (extras != null) {
        Object[] objArr = (Object[]) extras.get("pdus");
        SmsMessage smsMessage =  SmsMessage.createFromPdu((byte[]) objArr[0]);
        if (objArr.length >= 0) {
            String body = smsMessage.getMessageBody().toString().toUpperCase();
            if (body != null && body.startsWith("GETPOS")){sendPosition();}
        }
    }
}
\end{minted}
\caption{My Car Tracks decompiled (simplified)}
\end{listing*}

\section{Example of the approach}
\label{example}
\begin{figure*}[h]
    \centering
    \begin{subfigure}{.32\textwidth}
        \centering
\begin{minted}{java}
public void onReceive(Context c, Intent i) {
  SmsMessage sms = getSms(i);
//#sms
  String b = sms.getBody();
//#sms/#body
  String cmd = null;
//symbolic null value
  if(b.startsWith("!CMD:")){
//#sms/#body.startsWith("!CMD:")
    cmd = getCmdFromBody(b);
//symbolic string
    |\textcolor{red}{processCmd(cmd);}|
  }else{// do something else}
}
\end{minted}
        \caption{Symbolic execution (step B)}
        \label{example_se}
    \end{subfigure}
    \begin{subfigure}{.32\textwidth}
        \centering
\begin{minted}{java}
public void onReceive(Context c, Intent i) {
  SmsMessage sms = getSms(i);
  // -
  String b = sms.getBody();
  // -
  String cmd = null;
  // -
  if(b.startsWith("!CMD:")){
    cmd = getCmdFromBody(b);
    // $p$
    |\textcolor{red}{processCmd(cmd);}|
  }else{ // $\neg p$
  }
}
\end{minted}
        \caption{Path predicate recovery (step C)}
        \label{example_ppr}
    \end{subfigure}
    \begin{subfigure}{.32\textwidth}
        \centering
\begin{minted}{java}
public void onReceive(Context c, Intent i) {
  SmsMessage sms = getSms(i);
  String b = sms.getBody();
  String cmd = null;
  if(b.startsWith("!CMD:")){
// suspicious predicate
// code guarded by $p$
    cmd = getCmdFromBody(b);
    |\textcolor{red}{processCmd(cmd);}|
// sensitive method in
// inter-procedural call
  }else{// code guarded by $\neg p$
  }
}
\end{minted}
        \caption{Control dependency (steps D and E)}
        \label{example_cd}
    \end{subfigure}
    \caption{Example of the different steps of the analysis}
    \label{example_code}
\end{figure*}

\begin{multicols}{2}
We explain the process of the approach with Figure~\ref{example_code}.
First, we describe the symbolic execution step with the example of Listing~\ref{example_se}.
The first value modeled is at line 2, a new incoming SMS is being stored in variable $sms$ and is represented by the tag \texttt{\#sms}.
In line 4 the body of the SMS is retrieved, thus the instruction is tagged with \texttt{\#sms/\#body} as our symbolic execution engine recognizes it.
Some values cannot be resolved during this step, hence they are assigned symbolic values, e.g. at lines 6 and 10.
Those modeled values are useful to describe the semantic of the condition at line 8 which is represented by the tag \texttt{\#sms/\#body.startsWith("!CMD:")}.
This value will be used during the path predicate classification to qualify the suspiciousness of the condition.

Now that the analysis has tagged some interesting values, it retrieves the path predicate for each instruction in the code, as illustrated in Listing~\ref{example_ppr}.
This simple example shows that outside any conditions, intra-procedural instructions are not annotated with any logical formula (lines 3, 5 and 7).
However, instructions guarded by a condition are annotated with the predicate representing this condition.
Both branches are considered, that is why the instruction at line 9 is annotated with predicate $p$ and any instruction under the \textit{else} instruction is annotated with $\neg p$.
Those formulas are then subjected to minimization in order to rule out false dependencies.
It is now possible to classify predicates and to check if they guard sensitive operations.

The next phase, shown in Listing~\ref{example_cd}, allows classifying predicates thanks to the results of the previous symbolic execution.
Indeed, decisions about the suspiciousness are taken according to the results of the symbolic execution.
That is why the condition at line 5 is considered suspicious because the body of an incoming SMS is being matched against a hardcoded string.
Once the suspicious conditions memorized, the analysis retrieves the instructions guarded by those conditions thanks to the path predicate recovery step.
For each guarded instruction, the analysis checks if the instruction is a method call and, if it is the case, the called method is being matched against a list of sensitive methods.
If no match is found, as it is the case in~\ref{example_cd}, the inter-procedural mechanism takes place, meaning the analysis dives into application method calls to check if they use a sensitive method.
In this example, the \texttt{processCmd(String)} method at line 9 contains such a method call.
According to \ts{}'s approach, it is sufficient to qualify this sequence as a logic bomb.

\end{multicols}

\end{document}